\documentclass[twocolumn]{aastex62}
\newcommand{\IMBH}{\normalfont SDSS J0914+0853}
\newcommand{\BAYMAX}{{\tt BAYMAX}}
\shortauthors{Foord et al.}

\usepackage{amsmath,amssymb} 
\usepackage{graphicx,mathptmx} 
\usepackage{natbib} 

\newcommand{\beq}{
\begin{equation}
}
\newcommand{\eeq}{
\end{equation}
}

\newcommand{\beqa}{
\begin{eqnarray}
}
\newcommand{\eeqa}{
\end{eqnarray}
}

\begin{document}

\title{A Bayesian Analysis of \IMBH{}, a Low-Mass Dual AGN Candidate}

\author[0000-0002-1616-1701]{Adi Foord}
\affil{Department of Astronomy and Astrophysics, University of Michigan, Ann Arbor, MI 48109}

\author[0000-0002-1146-0198]{Kayhan G\"{u}ltekin}
\affil{Department of Astronomy and Astrophysics, University of Michigan, Ann Arbor, MI 48109}

\author[0000-0003-1621-9392]{Mark T. Reynolds}
\affil{Department of Astronomy and Astrophysics, University of Michigan, Ann Arbor, MI 48109}

\author[0000-0002-2397-206X]{Edmund Hodges-Kluck}
\affil{Department of Astronomy, University of Maryland, College Park, MD 20740}
\affil{Code 662, NASA Goddard Space Flight Center, Greenbelt, MD 20771}

\author[0000-0002-8294-9281]{Edward M. Cackett}
\affil{ Department of Physics and Astronomy, Wayne State University, Detroit, MI, 48202}

\author[0000-0001-8627-4907]{Julia M. Comerford}
\affil{Department of Astrophysical and Planetary Sciences, University of Colorado, Boulder, CO, 80309}

\author{Ashley L. King}
\affil{Department of Physics, Stanford University, Stanford, CA 94305}

\author{Jon M. Miller}
\affil{Department of Astronomy and Astrophysics, University of Michigan, Ann Arbor, MI 48109}

\author[0000-0001-8557-2822]{Jessie C. Runnoe}
\affil{Department of Astronomy and Astrophysics, University of Michigan, Ann Arbor, MI 48109}



\begin{abstract}
We present the first results from \BAYMAX{} (Bayesian AnalYsis of Multiple AGN in X-rays), a tool that uses a Bayesian framework to quantitatively evaluate whether a given \emph{Chandra} observation is more likely a single or dual point source.  Although the most robust method of determining the presence of dual AGNs is to use X-ray observations, only sources that are widely separated relative to the instrument PSF are easy to identify. It becomes increasingly difficult to distinguish dual AGNs from single AGNs when the separation is on the order of \emph{Chandra}'s angular resolution ($<1\arcsec$). Using likelihood models for single and dual point sources, \BAYMAX{} quantitatively evaluates the likelihood of an AGN for a given source. Specifically, we present results from \BAYMAX{} analyzing the lowest-mass dual AGN candidate to date, \IMBH{}, where archival \emph{Chandra} data shows a possible secondary AGN $\sim0\farcs{3}$ from the primary.  Analyzing a new 50 ks \emph{Chandra} observation, results from \BAYMAX{} shows that \IMBH{} is most likely a single AGN with a Bayes factor of 13.5 in favor of a single point source model.  Further, posterior distributions from the dual point source model are consistent with emission from a single AGN.  We find the probability of \IMBH{} being a dual AGN system with a flux ratio $f>0.3$ and separation $r>0\farcs{3}$ to be very low.  Overall, \BAYMAX{} will be an important tool for correctly classifying candidate dual AGNs in the literature, and studying the dual AGN population where past spatial resolution limits have prevented systematic analyses.
\end{abstract}

\section{Introduction}
\label{sec:intro}
\par Given that almost all massive galaxies are thought to harbor nuclear supermassive black holes (SMBH; \citealt{Kormendy&Richstone1995}) and that classical heirarchical galaxy evolution predicts that later stages of galaxy evolution are governed by mergers (e.g., \citealt{WhiteandReese1978}), galaxy mergers provide a favorable environment for the assembly of active galactic nuclei (AGNs) pairs \citep{Volonteri2003}.  The role galaxy mergers play in triggering AGN and/or AGN pairs remains unclear (e.g. \citealt{Hopkins&Quataert2010, Kocevski2012, Schawinski2012, Hayward2014, Villforth2014, Villforth2017, Capelo2015}), however both observations and simulations agree that AGN activity should increase with decreasing galaxy separation (e.g. \citealt{Koss2012, Blecha2013, Ellison2013, Goulding2017, Capelo2017, Barrows2017}).  
\par ``Dual AGNs" are usually defined as a pair of AGN in a single galaxy or merging system (with typical separations of $\sim$1 kpc), while a ``binary AGN" is a pair of AGNs that are gravitationally bound with typical separations $\lessapprox$100 pc (see \citealt{Begelman1980} for a summary of the main merging phases of SMBHs).  Understanding the specific environments where dual AGNs occur provides important clues about black hole growth during the merging process.  Additionally, as progenitors to SMBH-binary mergers, the rate of dual AGNs is intimately tied to gravitational wave events detectable by pulsar timing arrays and space-based interferometry (see \citealt{Mingarelli2019} and references within). Thus, dual AGNs offer a critical way to observe the link between galaxy mergers, SMBH accretion, and SMBH mergers.
\par  The frequency of galaxy mergers in our observable universe implies that dual AGN should be relatively common.  In particular, assuming a dynamical friction timescale of $\sim$1 Gyr we expect the galaxy merger fraction with separations $\le$ 1 kpc by $z=0.1$ to be between $\sim$6--10\% \citep{Hopkins2010}.  However, these estimated merger fractions don't take into account the AGN duty cycle, and observations of nearby AGN have shown that at separations $\le$ 1 kpc the fraction of \emph{dual AGN} may be much higher \citep{Barrows2017}.  Yet, very few dual AGN with separations $<$ 1 kpc have been confirmed.  Such systems become difficult to resolve with \emph{Chandra} beyond $z\ge0.05$, where separations on the order of 1 kpc approach \emph{Chandra}'s angular resolution (where the half-power diameter is $\sim$0\farcs{8} at $\sim$1 keV).  For example, the closest dual AGN candidate identified using two resolved point sources with \emph{Chandra} is NGC 3393 \citep{Fabbiano2011} with a projected separation of $\sim$150 pc ($\sim$0\farcs6; however see \citealt{Koss2015} for a critical analysis of the X-ray emission). Thus, many indirect detection techniques have been developed to search for evidence of dual and binary AGNs, primarily relying on optical spectroscopy and photometry.
\par Perhaps the most popular method of finding dual AGN candidates is via double-peaked narrow line emission regions (e.g., \citealt{Zhou2004, Gerke2007, Comerford2009, Liu2010, Fu2012, Comerford2012, Comerford2013}).  Double-peaked narrow lines can be a result of a dual AGN system during the period of the merger when their narrow line regions (NLR) are well separated in velocity.  The system can display two sets of narrow line emission regions, such as [O~III], where the separation and width of each peak will depend on parameters such as the distance between the two AGN.  However the optical regime alone is insufficient in confirming a dual AGN candidate because of ambiguity in interpretation of the observed double-peaked narrow line regions.  For example, bipolar outflows and rotating disks can also can produce the double-peaked emission feature (see, e.g., \citealt{Greene&Ho2005a, Rosario2010, MullerSanchez2011, Smith2012,Nevin2016}). Indeed, follow-up observations using high-resolution imaging and spatially resolved spectroscopy have found that many double-peaked dual AGN candidates are most likely single AGN \citep{Fu2012, Shen2011, Comerford2015}.  Dual or binary AGN candidates can be confirmed using high resolution radio imaging (see \citealt{Rodriguez2006,Fu2015, MullerSanchez2015, Kharb2017}); however an absence of radio emission does not necessarily mean an absence of AGNs (as only $\sim10\%$ of AGN are radio-loud), while a detection of two radio nuclei can have multiple physical explanations (such as star-forming nuclei). Nuclei can only be classified as AGN at radio frequencies if they are compact and have flat or inverted spectral indices (see, e.g., \citealt{Burke-Spolaor2011,Hovatta2014}).
\subsection{X-ray observations of dual AGN candidates}
\par  The most robust method of confirming the presence of dual AGNs is to use X-ray observations.  Due to the relatively few possible origins of emission above $10^{40}$ erg s$^{-1}$ besides accretion onto a SMBH \citep{Lehmer2010}, X-rays are one of the most direct methods of finding black holes, especially with \emph{Chandra}'s superb angular resolution.  Unlike the optical regime, X-rays are less sensitive to absorption from the dusty environments of merger remnants.  Currently, many analyses searching for dual AGN candidates using \emph{Chandra} observations implement the Energy-Dependent Subpixel Event Respositioning (EDSER) algorithm \citep{Li2004}.  EDSER improves the angular resolution of \emph{Chandra}'s Advanced CCD Imaging Spectrometer (ACIS) by reducing photon impact position uncertainties to subpixel accuracy, and in combination with \emph{Chandra}'s dithering can resolve sub-pixel structure down to the limit of the \textit{Chandra} High Resolution Mirror Assembly.  However, thus far it has only been used to make images and qualitatively analyze them for dual point sources. In the absence of corroborating evidence from other data, the reliance on visual interpretation of dual AGNs with separations comparable to \emph{Chandra}'s resolution leads to both false negatives and false positives.  This issue is worse in the low-count regime ($<$ 200 counts), where even dual AGN with larger separations ($>0\farcs{5}$) but low flux ratios are not clearly distinct.
\par We have developed a {\tt PYTHON} tool \BAYMAX{} (\textbf{B}ayesian \textbf{A}nal\textbf{Y}sis of \textbf{A}GNs in \textbf{X}-rays) that allows for a quantitative and rigorous analysis of whether a source in a given \emph{Chandra} observation is more likely composed of one or two point sources. This is done by taking calibrated events from a \emph{Chandra} observation and comparing them to the expected distribution of counts for single or double source models.  The main component of \BAYMAX{} is the calculation of Bayes factor, which represents the ratio of the plausibility of observed data, given two different models. Values $>1$ or $<1$ signify which model is more likely (see Section~\ref{sec:methods} for explicit details).  Further, \BAYMAX{} returns the maximum likelihood values for the parameters of each model.  In this paper we introduce our tool \BAYMAX{} and present its analysis on the \emph{Chandra} observations of the lowest-mass dual AGN candidate \IMBH{}.  Here we specifically highlight \BAYMAX{}'s capabilities with respect to the \emph{Chandra} observations of \IMBH{}.  We are using a subset of \BAYMAX{}'s full capabilities, i.e., analyzing an on-axis source, assuming identical spectra for both the primary and secondary AGN, and the background contribution is deemed negligible. As well, false positives are only analyzed for regions in parameter space (such as count number, separation, and flux ratio) that are specific to \IMBH{}.  Our following paper (Foord et al. 2019 (\emph{in prep})) will expands upon the explicit details of \BAYMAX{}, including its capabilities of correctly identifying dual AGN as a function of observed flux, angular separations, off-axis angle, and flux ratios.  In this paper, we restrict our discussion to \BAYMAX{}'s abilities on our observations of \IMBH{}.
\subsection{SDSS J0914+0853}
\par  \IMBH{} was originally identified by \cite{Greene&Ho2007} as one of $\sim$200 low-mass SMBH based on ``virial'' black hole mass estimates, where the velocity dispersion and radius of the broad line region (BLR) were estimated from H$\alpha$ emission line characteristics. The system is at $z=0.14$ ($D_{L} = 661$ Mpc and $D_{A} = 509$ Mpc for a $\Lambda$CDM universe, where $H_{0}=69.6$, $\Omega_{M}=0.286$, and $\Omega_{\Lambda}=0.714$) and is a low-mass ($M_{BH} = 10^{6.3} M_{\odot}$), low-luminosity AGN.  \IMBH{} was observed by \emph{Chandra} as part of a Cycle 13 program targeting low-mass AGNs (Proposal ID:13858, PI:G\"ultekin). These data were taken to investigate the fundamental plane in the low-mass regime and thus are on-axis \citep{Gultekin2014}.  Analyzing the 15 ks \emph{Chandra} data with EDSER, the archival \emph{Chandra} exposure shows a possible secondary source 0\farcs3 away from the primary. Possible contamination from an ultraluminous X-ray source (ULX) is very low; following the methodology in \cite{Foord2017} we calculate the number of expected ULXs with $L_\ge10^{41}$erg s$^{-1}$ to be $<10^{-3}$ within a radius of 0\farcs{3} from the center of the galaxy.  If the emission is found to most likely originate from two point sources, it will be the lowest-mass dual AGN discovered, and analysis of this system paves the way for a better understanding of the role of mergers and AGN activity in low-mass systems. In particular, dual AGNs in low-mass galaxies with low luminosities are the perfect testbed for discerning between competing models for the connection between galaxy mergers and AGN activity. It has been argued that mergers can trigger high-luminosity AGN but not low-luminosity AGN, which are triggered by stochastic processes \citep{Hopkins&Hernquist2009, Treister2012}. A competing hypothesis is that there is no correlation between AGN luminosity and mergers (e.g.,\citealt{Villforth2014}). Since dual AGNs most likely arise from mergers, the presence of a low-luminosity dual AGN in \IMBH{} would show that low luminosity AGNs can arise from mergers. However, effects due to pileup and artifacts from the Point Spread Function (PSF) cannot be ruled out at a high statistical confidence for the low-count ($\sim$250 counts between $2$--$7$ keV) image. At 10\%, the pile-up fraction is relatively small, but combined with asymmetries in the \emph{Chandra} PSF (\citealt{Juda&Karovska2010}), it could produce a spurious dual AGN signature. Thus, a statistical analysis is necessary before a discovery can be confirmed.
\par We aim to unambigiously determine the true nature of  \IMBH{}. As stated above, the existing \emph{Chandra} data cannot do this because (i) the pile-up introduces systematic uncertainties in the EDSER processing, (ii) the existing exposure is relatively shallow (15 ks), and (iii) potential PSF artifacts can produce spurious dual AGN signatures.  To help determine the true nature of \IMBH{}, we received a new observation (Proposal ID:19464, PI:G\"ultekin) that addresses all three of the above points. In particular, the observation (i) uses the shortest possible frame time with a subarray, thereby eliminating pileup, (ii) goes 3$\times$ deeper with a 50 ks exposure, and (iii) uses a substantially different roll angle so that any PSF artifacts will not appear in the same location on the sky.  With a total of $\sim$723 counts between $2$--$7$ keV (combining both datasets), \BAYMAX{} is able to statistically analyze the likelihood that \IMBH{} is a dual AGN for separations $>0\farcs3$ and flux ratios $>0.1$.
\par The remainder of the paper is organized into 5 sections. Section 2 introduces Bayesian inference, focusing on the specific components of Bayes factor and how \BAYMAX{} calculates the likelihood and prior densities. In Section 3 we analyze the \emph{Chandra} observations of \IMBH{}, including both a photometric and spectral analysis. In Section 4 we present our results when running \BAYMAX{} on the \emph{Chandra} observations of \IMBH{}, and in Section 5 we discuss the sensitivity and limitations of \BAYMAX{} across parameter space and how they affect our results. Lastly, we summarize our findings in Section 6.  Please see Table~\ref{tab:Symbols} for a list of symbols used throughout this paper.
\begin{table*}[t]
\begin{center}
\caption{Symbols}
\label{tab:Symbols}
\small
\begin{tabular*}{0.6\textwidth
}{l@{\extracolsep{\fill}}l}
	\hline
	\hline
	\multicolumn{1}{c}{Symbol} & \multicolumn{1}{c}{Definition} \\
	\multicolumn{1}{c}{(1)} & \multicolumn{1}{c}{(2)} \\
	\hline
	($x_{i}, y_{i}$) & Sky coordinate of photon $i$ \\
	$E_{i}$ & Energy of photon $i$,  in keV \\
	$n$ & Total flux (counts) of given source \\
	$\mu$	& Central position of given source in sky coordinates (2D; $\mu = [\mu_{x}, \mu_{y}]$)	\\
	$k$ & Number of \emph{Chandra} observations being modeled\\
    $\Delta x_{K}$   &  Translational astrometric shift in $x$ ($K=[1, \dots,  k-1]$) \\
    $\Delta y_{K}$   &  Translational astrometric shift in $y$ \\
    $\Delta r_{K}$  & Radial astrometric shift ($\Delta r_{K}=\sqrt{ (\Delta x_{K})^{2} + (\Delta y_{K})^{2}}$)		\\
    $\Delta\phi_{K}$  & Rotational astrometric shift 	\\

	$f$	& Flux ratio between secondary and primary source ($0<n_{S}$/$n_{P}<1$)	\\
	$M_{j}$ & Given model being analyzed by \BAYMAX{} \\
	$\theta_{j}$    & Parameter vector for $M_{j}$, i.e. [$\mu$, $f$, $\Delta x_{K}$, $\Delta y_{K}$, $\Delta\phi_{K}$] . \\
	\hline 
\end{tabular*}
\end{center}
Note. -- Columns: (1) Symbols used throughout the text; (2) Definitions. 
\end{table*}
\section{Methods}
\label{sec:methods}
\subsection{Bayesian Inference}
\BAYMAX{} is capable of statistically and quantitatively determining whether a given observation is better described by a model composed of one or two point sources based on a Bayesian framework. A Bayesian approach combines all available information (using prior distributions and likelihood models) to infer the unknown model parameters (posterior distributions).  Bayes Theorem implies: 
\begin{equation}
\underbrace{\frac{P(M_{2}\mid D)}{P(M_{1}\mid D)}}_\text{Posterior odds} = \underbrace{\frac{P(D\mid M_{2})}{P(D\mid M_{1})}}_\text{Bayes factor} \times \underbrace{\frac{P(M_{2})}{P(M_{1})}}_\text{Prior odds},
\end{equation}
where the posterior odds represents the ratio of the dual point source model ($M_{2}$) vs. the single point source model ($M_{1}$) given the data $D$; the Bayes factor ($BF$) quantifies the evidence of the data for $M_{2}$ vs. $M_{1}$, and the prior odds represents the prior probability ratio of $M_{2}$ vs. $M_{1}$. Specifically, the Bayes factor is the ratio of the marginal likelihoods: 
\begin{equation}
BF = \frac{\int P(D\mid\theta_{2},M_{2}) P(\theta_{2}\mid M_{2}) d\theta_{2}}{\int P(D\mid\theta_{1},M_{1}) P(\theta_{1}\mid M_{1}) d\theta_{1}},
\end{equation}
representing the ratio of the plausibility of observed data $D$, given two different models, and parameterized by the parameter vectors $\theta_{2}$ and $\theta_{1}$. Values $>$1 or $<$1 signify whether $M_{2}$ or $M_{1}$ is more likely (see \citealt{Jeffreys1935} and \citealt{Kass&Raftery1995} for the historic interpretations of the strength of a $BF$ value; we analyze our data to define a ``strong" $BF$ value in Section \ref{sec:discussion}.) In this paper, we assume that $M_{2}$ and $M_{1}$ are equally probable, so that $P(M_{2}) = P(M_{1}) = 0.5$ and the Bayes factor directly represents the posterior odds.  Thus calculating the Bayes factor can be broken into two components\,---\,the likelihood density, $P(D\mid\theta_{j},M_{j})$, and the prior density, $P(\theta_{j}\mid M_{j})$.  
\subsection{Data Structure and Modeling the PSF}
In this section we will focus on the likelihood density implemented in \BAYMAX{}. Each reprocessed \emph{Chandra} level-2 event file tabulates the directional coordinates ($x_{i}$, $y_{i}$) and energy $E_{i}$ for each detected photon, where $i$ indexes each detected photon (See Table~\ref{tab:Symbols} for a summary of notation).  The detector itself records the pulse height amplitude (PHA) of each event, which is roughly proportional to the energy of the incoming photon.  In the reprocessed files, the energy $E_{i}$ is calculated from the event's PHA value, using the appropriate gain table.  Thus, \BAYMAX{} takes calibrated events ($x_{i}$, $y_{i}$, $E_{i}$) from reprocessed \emph{Chandra} observations and compares them to new simulations based on single and dual point source models.
\par We characterize the properties of the \emph{Chandra} PSF by simulating the PSF of the optics from the High Resolution Mirror Assembly (HRMA) via ray tracing simulations.  The two primary methods to simulate the HRMA PSF are {\tt SAOTrace}\footnote{http://cxc.harvard.edu/cal/Hrma/Raytrace/SAOTrace.html} and the Model of AXAF Response to X-rays ({\tt MARX}, \citealt{Davis2012}). While the {\tt MARX} model uses a slightly simplified (and faster) description of the HRMA, differences between {\tt SAOTrace} and {\tt MARX} simulations are minimal for on-axis simulations. For our PSF analysis below, we find consistent parametric fits between an {\tt SaoTrace} generated PSF and one generated by {\tt MARX} -- in particular the root-mean-square error between the two fits is on the order of $\sim0.1\%$ 
\par Thus, our Likelihood models for single and dual point sources are created by parametrically modeling the \emph{Chandra} PSF using high count simulations created by {\tt MARX}-5.3.3. To translate the PSF model to an event file, the HRMA ray tracing simulations are projected on to the detector-plane via {\tt MARX}. Ray tracing simulations generated by both {\tt MARX} and {\tt SAOTrace} will have roughly the correct total intensity, but small deviations in the overall shape. Specifically regarding {\tt MARX} --  the PSF wings are broader than observations while the PSF core is narrower than observed \citep{Primini2011}. These discrepancies can be reduced by blurring the PSF when projecting it to the detector-plane via the {\tt AspectBlur} parameter. This parameter is used to account for the uncertainty in the determination of the aspect solution (such as effects from pixel quantization and pixel randomization), as well as the uncertainty in the instrument and dither models within {\tt MARX}.  The best value should be considered carefully for each unique observation\footnote{http://cxc.harvard.edu/ciao/why/aspectblur.html}.  For {\tt MARX} generated simulations on ACIS-S, we expect the {\tt AspectBlur} parameter to have values between $0\farcs25-0\farcs28$.  For our PSF analysis we set {\tt AspectBlur} to $0\farcs28$. We note that value used for {\tt AspectBlur} does not represent the accuracy at which we can centroid.
\par  For a given observation, a user-defined source model is input to {\tt MARX} to generate X-ray photons incident from a single point source centered on the observed central position of the AGN ($\mu_{obs}$, defined as the coordinates where the hard X-ray emission from the AGN is estimated to peak).  Because we do not model the spectral parameters of the system (see Section~\ref{sec:intro}), we are only interested in modeling the spatial distribution of a photon due to its energy $E_{i}$ and our PSF does not depend on the spectral shape of our model.  Each simulation uses the observation-specific detector position ({\tt RA\_Nom, Dec\_Nom, Roll\_Nom}) and start time ({\tt TSTART}).  We set the number of generated rays ({\tt NumRays}) to $1\times10^{6}$ and the read-out strip is excluded by setting the parameter {\tt ACIS\_Frame\_Transfer\_Time} to 0.  
\par We model the PSF as a summation of 2D Gaussians, where the amplitude and standard deviation of each Gaussian is energy-dependent.  In general, the PSF may be any function which is unique to a given observation and can be quickly evaluated.  For both the 15 ks and 50 ks observation, we fit a variety of possible functions to the PSF.  Using the Bayesian Information Criterion as a diagnostic for model comparison, we find that a summation of three circular concentric 2D Gaussians yields the best-fit for both \emph{Chandra} observations (specifically, we find the PSF wings are best-modeled by the broadness of a Gaussian component versus the addition of a Lorentzian component).  We model the PSF for each observation individually, however we find that the best-fit parameters for each PSF model are consistent with one-another within the $1\sigma$ error bars (which is not surprising, given that both sources were observed on-axis for ACIS-S albeit in different \emph{Chandra} cycles).  
\par Each photon is assumed to originate from a single or dual point source system.  For example, for a single point source, the probability that a photon observed at location $x_{i}$,$y_{i}$ on the sky with energy $E_{i}$ is described by the PSF centered at $\mu$ is $P(x_{i},y_{i}\mid \mu, E_{i})$, i.e., the energy dependent PSF.  For $n$ total events, the total probability is the product of the probability for each detected photon, i.e., the likelihood density is:
\begin{equation}
    \begin{split}
    \mathcal{L}&=P(x,y\mid \mu,E) = \prod_{i=1}^{n} P(x_{i},y_{i}\mid \mu,E_{i}) \\
    & = \prod_{i=1}^{n} \frac{ M_{1,i}(\theta_{1})^{D_{i}}}{D_{i}!} \exp(-M_{1,i}(\theta_{1})),
    \end{split}
\end{equation}
where we use the Poisson likelihood, appropriate given that \emph{Chandra} registers each event individually. Here, $M_{1, i}(\theta_{1})$ is the probability for event $i$ given our PSF model, and $D_{i}$ is the data value for event $i$.  For a dual point source the total probability is $P(x, y \mid \mu_{P},\mu_{S}, E, n_{S}/n_{P})$, where $\mu_{P}$ and $\mu_{S}$ represent the location of the primary and secondary AGN.  The ratio of the fluxes (or, total counts) between the secondary and primary is represented by $n_{S}/n_{P} = f$, where $0\le f\le1$.  We note that our analysis on \IMBH{} does not include fitting for the spectral models.  Using the archival data we find consistent hardness ratios between the candidate primary and secondary AGN, where we use circular and non-overlapping apertures centered on their apparent locations.  Thus, we assume that the spectra are the same spectral shape as that for the entire system, but with different normalizations.  Future analyses with \BAYMAX{} will include fitting for different spectral shapes. 
\par  Because Bayes factor represents the \emph{ratio} of likelihood densities, and we use the same data across both models, our calculations become simplified. We are left with: 
\begin{equation}
    \mbox{ln} \mathcal{L} = \sum_{i=1}^{n} D_{i}\ \mbox{ln}M_{j, i}(\theta_{j}) + \mbox{constant}, 
\end{equation}
where $M_{j, i}(\theta_{j})$ is calculated for either a single ($j=1$) or dual ($j=2$) point source model via our parametrically-fit PSF, for each detected event $i$.
\subsection{Prior Distributions}
\BAYMAX{} requires user input regarding (i) the number of datasets and (ii) the prior distributions for each parameter.  Regarding point (i), \IMBH{} has $k=2$ observations and thus the parameter vector $\theta_{1}=[\mu, \Delta x_{1}, \Delta y_{1}]$ while $\theta_{2}=[\mu_{{P}}$, $\mu_{{S}}, {\log{f}},  \Delta x_{1}, \Delta y_{1}]$.  Here, $\mu = (\mu_{x}, \mu_{y})$ is the central sky $x$,$y$ positions of the AGN; $\Delta x_{1}$ and $\Delta y_{1}$ account for the translational components of the relative astrometric registration for the $k-1$ observation; and $\log{f}$ is the log of the flux ratio where $f=n_{S}/n_{P}$. The relative astrometric registration adds an uncertainty that must be taken into account in order to avoid spurious dual AGN signals that can be generated from slight mismatches between two or more observations. We take this into account by including the astrometric registration of multiple observations as a set of parameters to be marginalized over. For \IMBH{}, we  find that the rotational component of the relative astrometric registration is expected to be very small ($\Delta \phi_{1}<1^{\circ}$) and including the parameter does not affect our results.  Thus, we only include $\Delta x_{1}$ and $\Delta y_{1}$, which are analyzed for the shallower observation (i.e., relative to the 50 ks exposure). 
\par Regarding point (ii), \BAYMAX{} can incorporate any user-defined function to describe the prior distributions for each parameter.  For \IMBH{}, the prior distributions of $\mu$ for both $M_{1}$ and $M_{2}$ are described by a continuous uniform distribution\footnote{The continuous uniform distribution $\mathcal{U}(a,b)$ is a probability distribution where all values between the minimum $a$ and maximum $b$ are equally probable.}:
\begin{equation}
    	\mu = {\mathcal {U}}(a,b),
\end{equation}
where we constrain all $\mu$ values to be between $a=\mu_{obs}-2$ and $b=\mu_{obs}+2$.  Thus, the 2D parameter space for possible $\mu_{x}$ and $\mu_{y}$ is a 4$\times$4 sky-pixel box ($\approx$ 1.98\arcsec $\times$1.98\arcsec) centered on the observed central X-ray coordinates of \IMBH. Further, the prior distributions of $\Delta x_{1}$ and $\Delta y_{1}$ are also described by a uniform distribution with $a=\delta \mu_{obs}-3$ and $b=\delta \mu_{obs}+3$, where $\delta \mu$ represents the difference between the observed central X-ray coordinates of the two observations (in practice, $\delta \mu$ is expected to be small; however because the most recent observation of \IMBH{} was taken in a subarray mode, the difference between the aimpoints of the two observations is $\approx$15 sky pixels).  For $M_{2}$, the prior distribution for $\log{f}$ is also described by a uniform distribution (and thus $f$ is described by a \emph{log uniform} distribution), where $a=-2$ and $b=0$.  The range for the prior distribution of $\log{f}$ covers possible values expected for ``major mergers'' (with mass ratios $>1/3$), while accounting for a large range of possible Eddington fractions between the two black holes.  In general, informative priors can be incorporated if prior information is available. For example, we might set the prior distributions of $\mu_{P}$ and $\mu_{S}$ to Gaussian distributions centered on coordinates that are better constrained by other observations (such as spectroastrometric [O~III] observations, or complementary IR photometry). 
\subsection{Calculation of Bayes Factor}
Bayesian inference can be divided into two categories: model selection and parameter estimation. In this section we review how we address each component in \BAYMAX{}. \\
\par Computing the marginal likelihood is challenging, as it involves a multi-dimensional integration over all of parameter space. Only over the last $\sim$20 years have the advances in computational power allowed Bayesian inference to become a more common technique for model selection.  In addition, general numerical methods based on Markov chain Monte Carlo (MCMC; e.g., see \citealt{Metropolis1953}) have been developed, allowing one to conduct Bayesian inferences in an efficient manner, with few constraints on dimensionality or analytical integrability.
\par To calculate the marginal likelihood, \BAYMAX{} implements a sampling technique called nested sampling \citep{Skilling2004}.  In nested sampling, the marginal likelihood is rebranded as the ``Bayesian evidence", denoted by $Z$.  Here, $Z = \int P(D\mid\theta_{j},M_{j}) P(\theta_{j}\mid M_{j}) d\theta_{j}$. Nested sampling transforms the multi-dimensional integral to a one dimensional integral by introducing the \emph{prior mass} $X$, defined as $X=\int_{\mathcal{L}(\theta)>\lambda} P(\theta_{j}\mid M_{j}) d\theta_{j}$; here the integral extends over the regions of parameter space contained within the iso-likelihood contour $\mathcal{L}(\theta)=\lambda$ and at any given time has a value $0<X<1.$ For a step-by-step explanation of nested sampling we refer the reader to \cite{Skilling2004}, \cite{Shaw2007}, \cite{Feroz&Hobson2008}, and \cite{Feroz2009}.  A direct effect of the nested sampling methodology is sparsely sampling in low likelihood regions and densely sampling where the likelihood is high.   To calculate the evidence, \BAYMAX{} uses the {\tt PYTHON} package {\tt nestle}\footnote{https://github.com/kbarbary/nestle}.  The package provides a pure-{\tt PYTHON} implementation of nested sampling, where prior mass space can be sampled via different techniques.  In particular, we use multi-ellipsoidal sampling (by setting {\tt method}=`multi'; see \citealt{Mukherjee2006, Shaw2007, Feroz&Hobson2008}).
\par The two parameters that affect the accuracy of $Z$ are the number of ``active points" and the stopping criterion dlog$Z$. The number of active points represent how many points in prior mass space one is sampling at a given time (roughly analogous to the number of walkers in an MCMC run).  The stopping criterion determines when the nested sampling loop terminates\,---\,when the current largest sampled likelihood does not increase by more than the stopping criterion value, the sampling will end\footnote{Specifically, at a given iteration $i$ where the current evidence is $Z_{i}$ and the estimated remaining evidence in the likelihood landscape is $Z_{est}$, if $\mbox{log}(Z_{i} + Z_{est}) - \mbox{log}(Z_{i}) < \mbox{dlog}Z$, the sampling will terminate}.  For our analysis of \IMBH{}, we use 500 active points (generally, a lower limit on the number of active points is 2$N_{dim}$, and due to using the multi-ellipsoidal method we add additional points to characterize each mode well) and set the stopping criterion dlog$Z=0.1$.  By continually increasing the number of active points and decreasing the stopping criterion, the estimated evidence and its accuracy should converge.  We find no significant difference in results when increasing our active points above 500 and using dlog$Z<0.1$, and conclude that these values have properly sampled the likelihood space.
\par Lastly, for parameter estimation \BAYMAX{} uses {\tt PyMC3} \citep{Salvatier2016}, which uses gradient-based MCMC methods for sampling.  Specifically, we use {\tt PyMC3}'s built in Hamiltonian Monte Carlo (HMC) sampling method.  HMC uses the gradient information from the likelihood to much more quickly converge than normal Metropolis-Hastings sampling.  In general, HMC is more powerful for high dimensionality and complex posterior distributions (see, e.g., \citealt{Betancourt2014}).
\par
\begin{figure*}
\centering
    \includegraphics[width=0.35\linewidth]{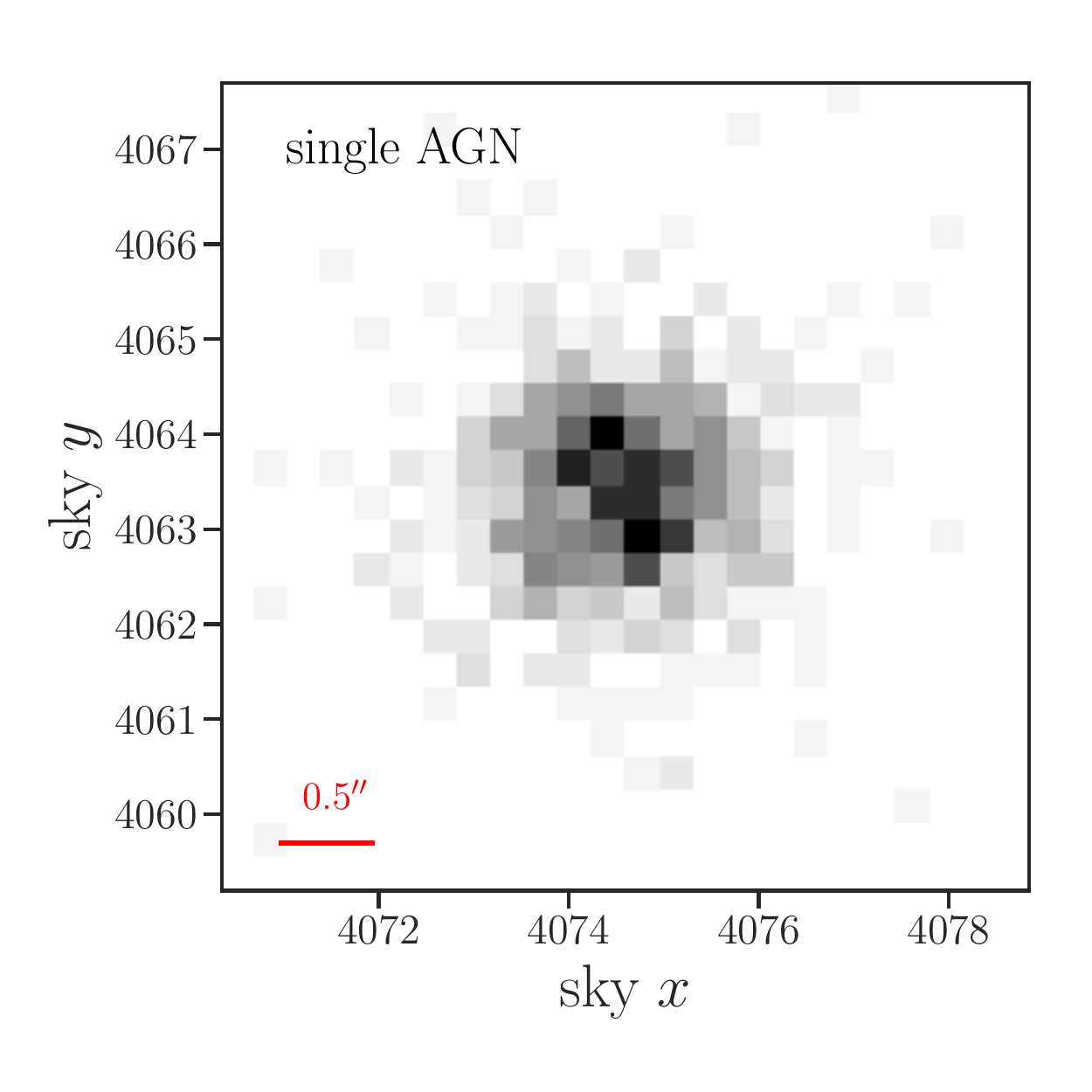}{\hskip 0pt plus 0.3fil minus 0pt}
    \includegraphics[width=0.355\linewidth]{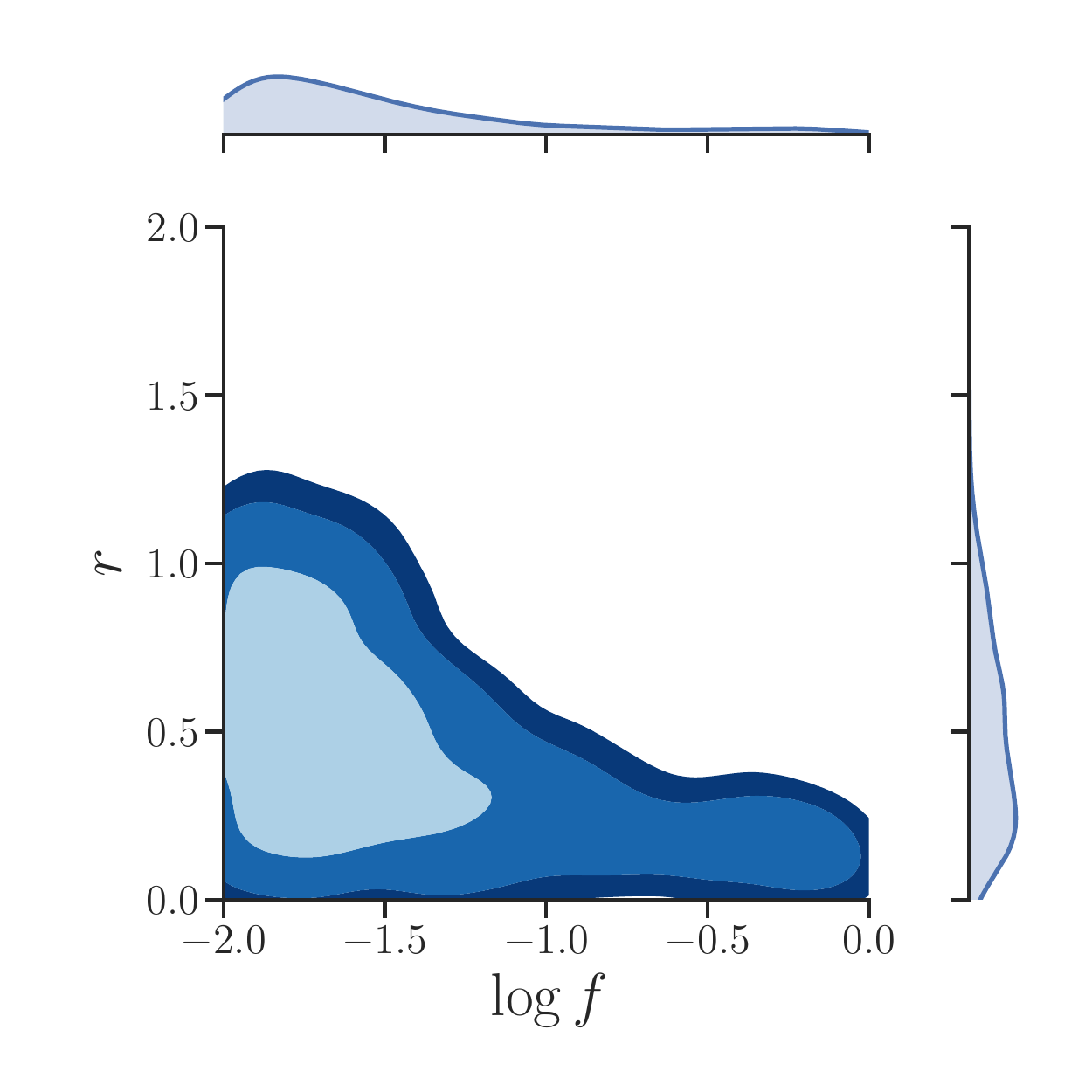}{\vskip 0.2cm plus 1fill}
    \includegraphics[width=0.35\linewidth]{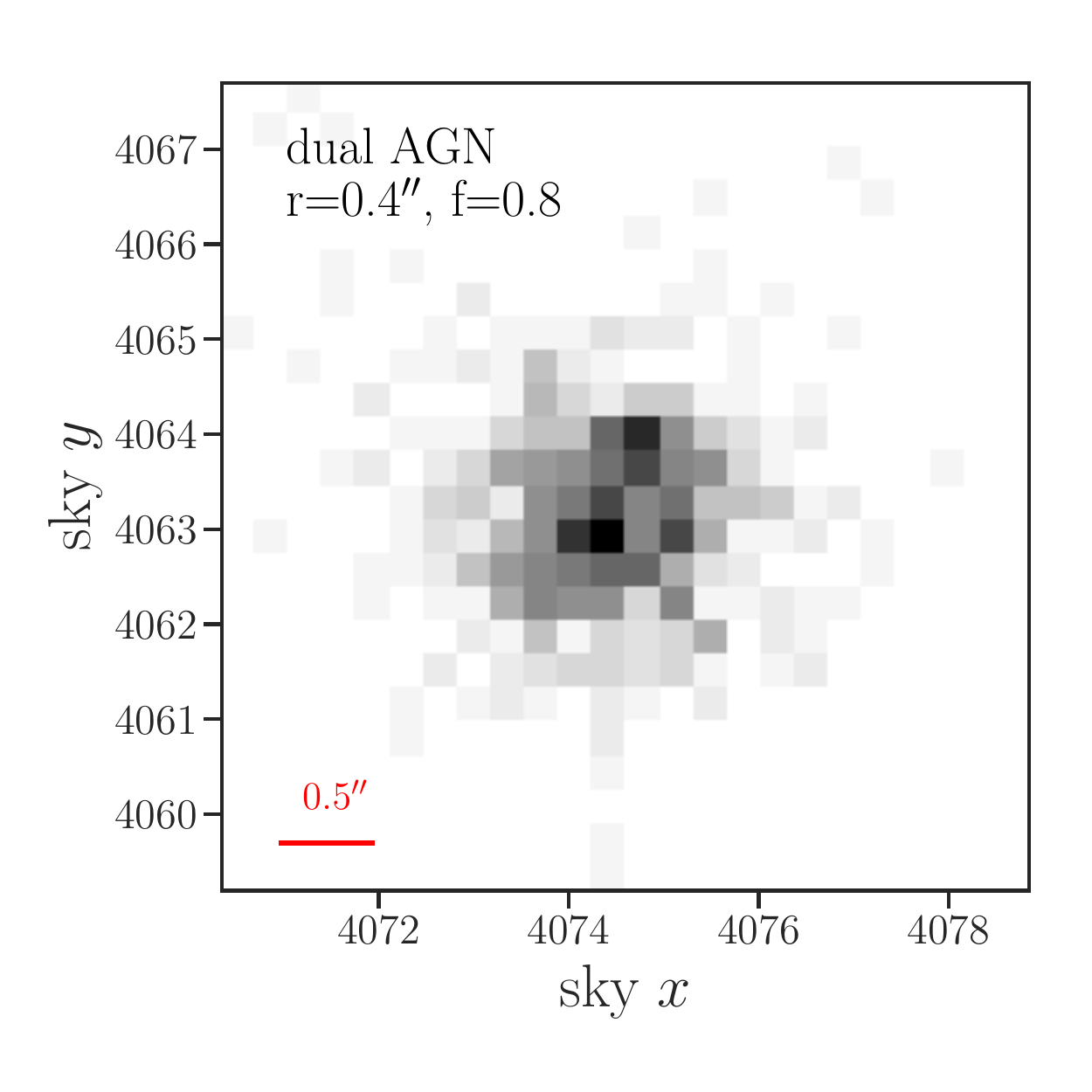}{\hskip 0pt plus 0.3fil minus 0pt}
    \includegraphics[width=0.355\linewidth]{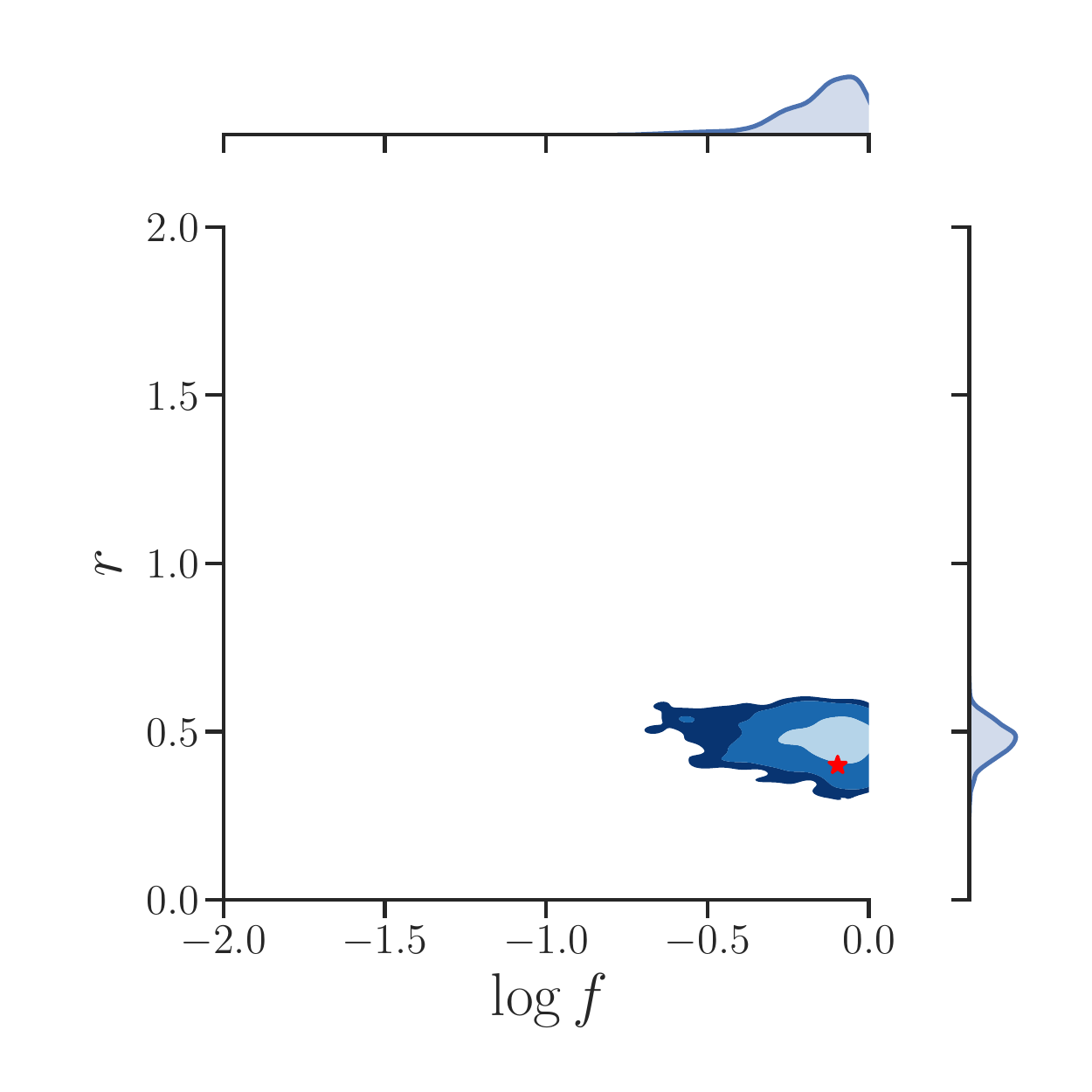}{\vskip 0.2cm plus 1fill}
    
\caption{\emph{Top}: A simulated single AGN (left) and the joint posterior distribution (right) for the separation ($r$, in arcsec)and flux ratio ($\log{f}$). The simulation has a total of $n=700$ counts between $2$--$7$ keV.  The simulations have been reprocessed using the Energy-Dependent Subpixel Event Repositioning algorithm (EDSER; \citealt{Li2004}), and binned by $2/3$ of the native pixel size. We do not include a background contribution from in these simulations.  Using \BAYMAX{}, we calculate a $BF$ strongly in favor of the single point source model. The joint posterior distribution is shown with the marginal distributions along the top and right border. 68\%, 95\% and 99.7\% confidence intervals are shown in blue contours. The separation and logarithm of the flux ratio are consistent with $0$ and $-2$ at the 99.7\% confidence level. We note that this particular joint-distribution shape is consistent with a single AGN, where at very large flux ratios the dual AGN candidate is likely to have $r=0$, and at very large separations the dual AGN candidate is likely to have $\log{f}=-2$. \emph{Bottom}: A simulated dual AGN (left) and the joint posterior distribution for the separation and flux ratio (right). The simulation has a separation $r=0\farcs4$ and $f=0.8$, and a total of $n=700$ 2-7 keV counts. It is difficult to tell whether the observation is composed of one or two point sources from the hard X-ray emission alone.  Using \BAYMAX{}, we calculate a $BF$ strongly in favor of the dual point source model. Further, using \BAYMAX{} we retrieve the correct separation and flux ratio values within the 68\% confidence level.}
\label{fig:BAYMAXsimulations}
\end{figure*}

 In Figure~\ref{fig:BAYMAXsimulations} we show the results when we analyze two simulations with \BAYMAX{}.  The simulations have been reprocessed using EDSER, and binned by $2/3$ of the native pixel size.  We simulate a single and a dual AGN system via {\tt MARX} using the same telescope configuration as our new 50 ks observation.  Both simulations have $n=700$ photons between $2$--$7$ keV, and the same $2$--$7$ keV spectrum as \IMBH{}.  We do not include a background contribution in our simulations (see Section~\ref{analysis}).  For the dual AGN simulation, each AGN has the same spectra but with normalizations such that the flux ratio $f = n_{S}/n_{P} = 0.8$ while the separation between the two AGN is 0\farcs4. As evident, it is difficult to visually distinguish whether a given simulation is actually composed of one or two sources. Using the methodology presented above, \BAYMAX{} favors the correct model for both simulations: for the single AGN simulation \BAYMAX{} estimates a $BF$ of $17\pm1.6$ in favor of the single point source model, while for dual AGN simulation \BAYMAX{} estimates a $BF$ of $25\pm1.5$ in favor of the dual point source model (error bars have been determined by running \BAYMAX{} multiple times on each simulation, see Section~\ref{sec:results}). Although the hard X-ray emission appears quite similar between the two simulations, the joint posterior distributions are significantly different from one another. Specifically, for the dual AGN simulation the joint posterior distribution is more tightly concentrated around the true values. \BAYMAX{} is able to recover the true separation and flux ratio within the 68\% credible interval. However, for the single AGN, the separation and flux ratio are consistent with 0 at the 99.7\% confidence level. We note that this particular joint-distribution shape (``L" shape) is consistent with a single AGN, where at very large flux ratios the dual AGN candidate is likely to have $r=0$, and at very large separations the dual AGN candidate is likely to have $\log{f}=-2$. More specifically, the dual point source model places each AGN at the same location and/or with arbitrarily low $f$, effectively consistent with one point source.

\begin{figure*}
\centering
    \includegraphics[width=0.3\linewidth]{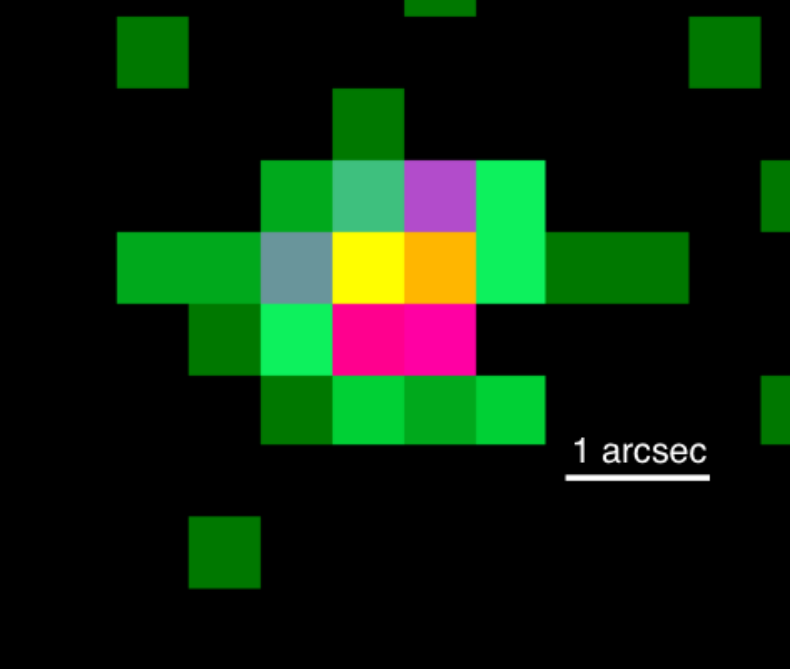}{\hskip 0pt plus 0.3fil minus 0pt}
    \includegraphics[width=0.307\linewidth]{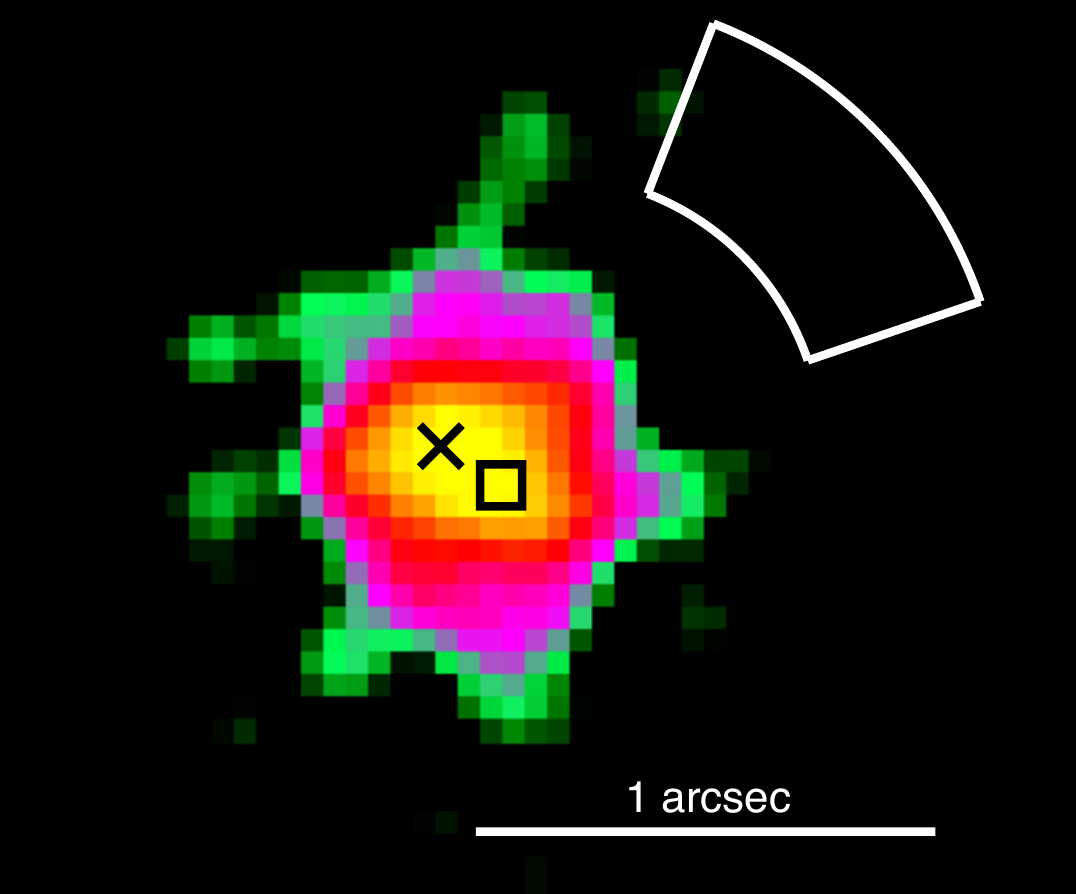}{\vskip 0.4cm plus 1fill}
    \includegraphics[width=0.3\linewidth]{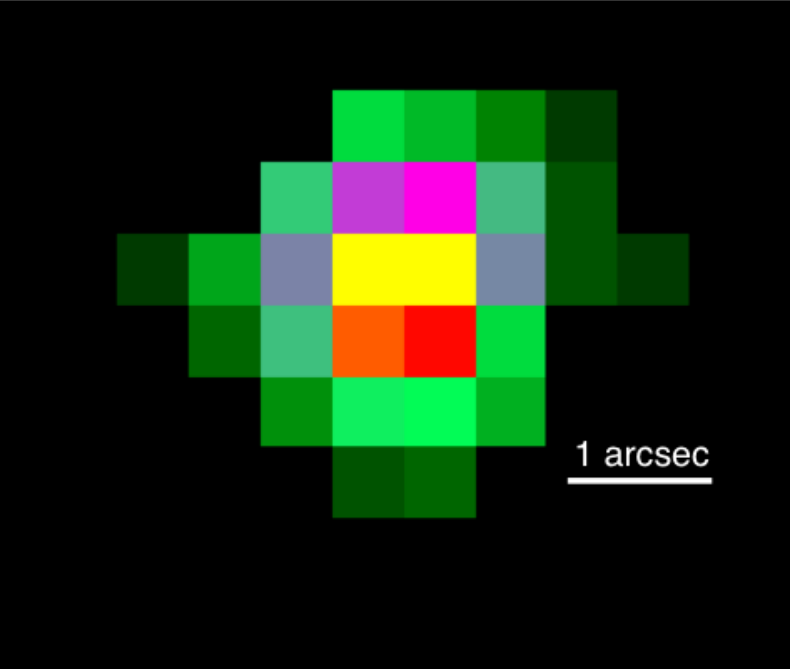}{\hskip 0pt plus 0.3fil minus 0pt}
    \includegraphics[width=0.307\linewidth]{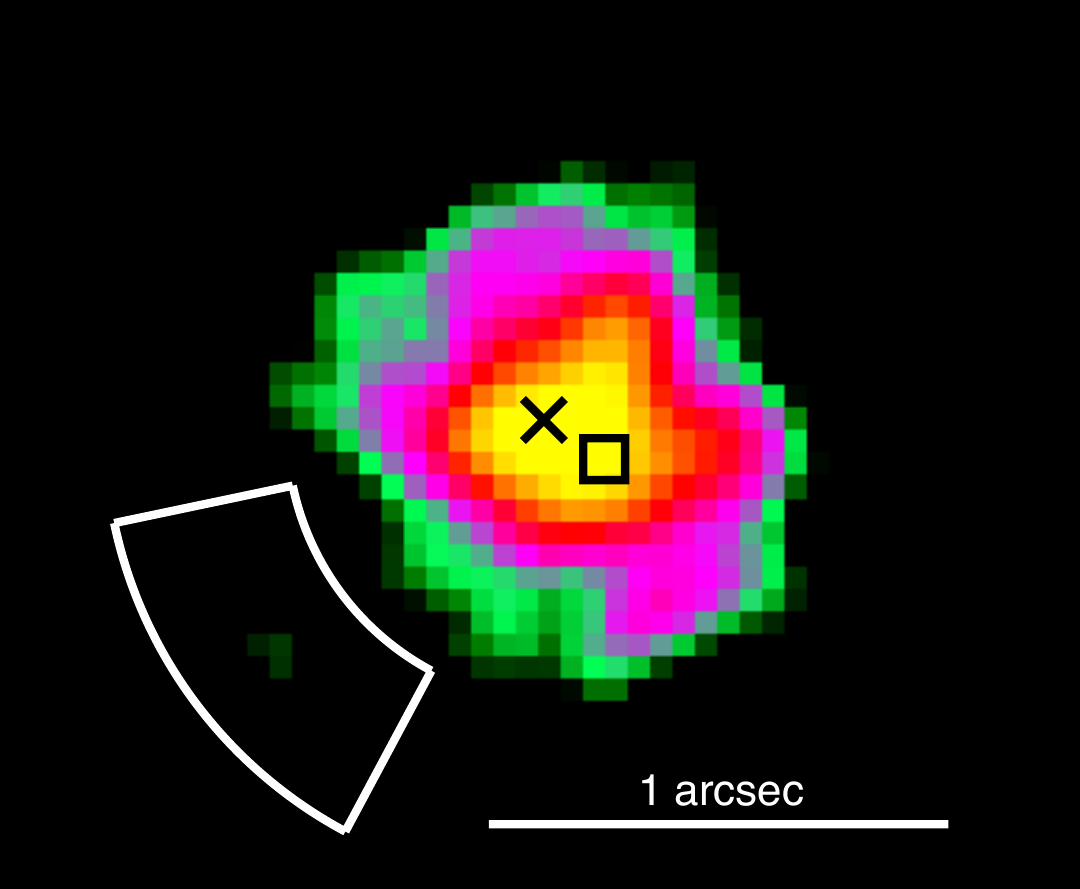}{\vskip 0.2cm plus 1fill}
    \vspace{\floatsep}

\caption{\emph{Chandra} images of \IMBH{}. \emph{Top}: $2$--$7$ keV raw (left) and smoothed (right) images of the 15 ks archival observation (Obs ID: 13858). The total number of $2$--$7$ keV counts shown is 257. The smoothed image has been reprocessed using the Energy-Dependent Subpixel Event Repositioning (EDSER; \citealt{Li2004}) algorithm, and binned by a tenth of the native pixel size. The location of the asymmetry in the \emph{Chandra} PSF is $\approx$ 0\farcs7 from the central position of the AGN, and is outlined by a white polygon. For both datasets, we mask the photons from this region before running \BAYMAX{}. There appear to be two regions of X-ray emission (denoted by a black "x" and a black square) separated by $\sim$0\farcs3. \emph{Bottom}: $2$--$7$ keV raw (left) and smoothed (right) image of our new 50 ks observation (Obs ID: 19464). The total number of $2$--$7$ keV counts is 484; the smoothed image has been reprocessed similarly to the archival dataset. We plot the spatial location of the primary (black "x") and secondary (black square) AGN, given the 15 ks observation. Although the archival dataset appears to have X-ray emission associated with two point sources, the new dataset has emission that more closely resembles a single point source.}
\label{fig:ChandraImages}
\end{figure*}

\section{Data Analysis}
\label{analysis}
\subsection{X-ray Data}
\IMBH{} was originally targeted to study low-mass AGNs and their relation to the plane of black hole accretion. The quasar was placed on the back illuminated S3 chip of the Advanced CCD Imaging Spectrometer (ACIS) detector, with an exposure time of 15 ks (Obs ID: 13858).  We received a new 50 ks exposure at a roll angle significantly different from the previous observation, and using the smallest subarray (1/8) on a single chip to get the shortest standard frame time (Obs ID: 19464). This was done to (1) place the PSF artifact in a different location, (2) avoid pileup, and (3) receive $\sim$ 2--3 times more counts. We re-reduced and re-analyzed the archival data to ensure a uniform analysis between the two datasets.
\par We follow a similar data reduction as described in previous X-ray studies analyzing AGN (e.g., \citealt{Foord2017, Foord2017b}), using the \emph{Chandra} Interactive Analysis of Observations ({\tt CIAO}) v4.8 \citep{Fruscione2006}.  Both datasets are analyzed with the energy-dependent sub-pixel event repositioning algorithm
(EDSER; \citealt{Li2004}), which can be included in the standard {\tt CIAO} reprocessing command {\tt chandra\_repro} with the parameter {\tt pix\_adj=EDSER}.  For each observation, we first evaluate the aspect solutions of the reprocessed level-2 event files to ensure the Kalman lock was stable at all times.  Further, we inspect the event detector coordinates as a function of time and find that they followed the instrument's dither pattern, indicating no aspect-correction based degradation of the PSF.
\par We then correct for astrometry, cross-matching the \emph{Chandra} detected point-like sources with the Sloan Digital Sky Survey Data Release 9 (SDSS DR9) catalog. The \emph{Chandra} sources used for cross-matching are detected by running {\tt wavdetect} on the reprocessed level-2 event file. We require each matched pair to be less than 2\arcsec~from one another and have a minimum of 3 matches.  The 15 ks observation meets the criterion for an astrometrical correction; we find 8 matches between the \emph{Chandra} observation and the SDSS DR9 catalog, resulting in a shift less than 0\farcs5. The 50 ks observation was taken in a subarray, and thus does not meet these criterion (however, because {\tt BAYMAX} takes into account astrometric shifts between observations this step will not affect our final results, see Section~\ref{sec:results}). Background flaring is deemed negligible as neither dataset contain intervals where the background rate is 3$\sigma$ above the mean level.
\par We then rerun {\tt wavdetect} on filtered 0.5 to 7 keV data to generate a list of X-ray point sources. We use wavelets of scales 1, 1.5, and 2.0 pixels using a 1.5 keV exposure map, and set the detection threshold significance to $10^{-6}$ (corresponding to one false detection over the entire S3 chip). We identify the quasar as an X-ray point source 0\farcs{4} (Obs ID: 13858) and 0\farcs{7} (Obs ID:19464) from the SDSS-listed optical center (2\arcsec corresponds to 95\% of the encircled energy radius at 1.5 keV for ACIS). Counts are extracted from a 2\arcsec\ radius circular region centered on the X-ray source center, where we use a source-free annulus with an inner radius of 20\arcsec\ and outer radius of 30\arcsec\ for the background extraction.  We compare the estimated background contribution from the datasets to the \emph{Chandra} blank-sky files.  Here, the blank-sky files are properly scaled in exposure time and have matching WCS coordinates, dimensions, and energies. We find consistent results, where within $2$--$7$ keV we expect $\lesssim$ 1 and 1.5 background counts within a 4$\times$4 sky-pixel box ($\approx$ 1.98\arcsec $\times$1.98\arcsec) centered on the quasar for the archival and new dataset, respectively. 
\par Our reduced data are shown in Figure~\ref{fig:ChandraImages}. Here, both exposures have been reprocessed using the EDSER algorithm and are binned by a tenth of the native pixel size.  The archival data appear to have sub-pixel structure, indicating a possible secondary AGN $\sim$ 0$\farcs$3 away from the primary, however our new observation is inconsistent with this picture. Although the X-ray emission may be slightly extended in the East-West direction (North is up, while East is left), we find minimal photometric evidence supporting the presence of a secondary AGN. 

\subsection{Spectral Fitting}
\label{sec:specfitting}
The quasar's net count rate and flux value are determined using XSPEC, version 12.9.0 \citep{Arnaud1996}. All errors evaluated in this section are done at the 95\% confidence level, unless otherwise stated, and error bars quoted are calculated with Monte Carlo Markov chains via the XSPEC tool {\tt chain}.  We implement the Cash statistic ({\tt cstat}; \citealt{Cash1979}) in order to best assess the quality of our model fits.  
\par
\begin{figure*}
\centering
\includegraphics[width=0.55\linewidth]{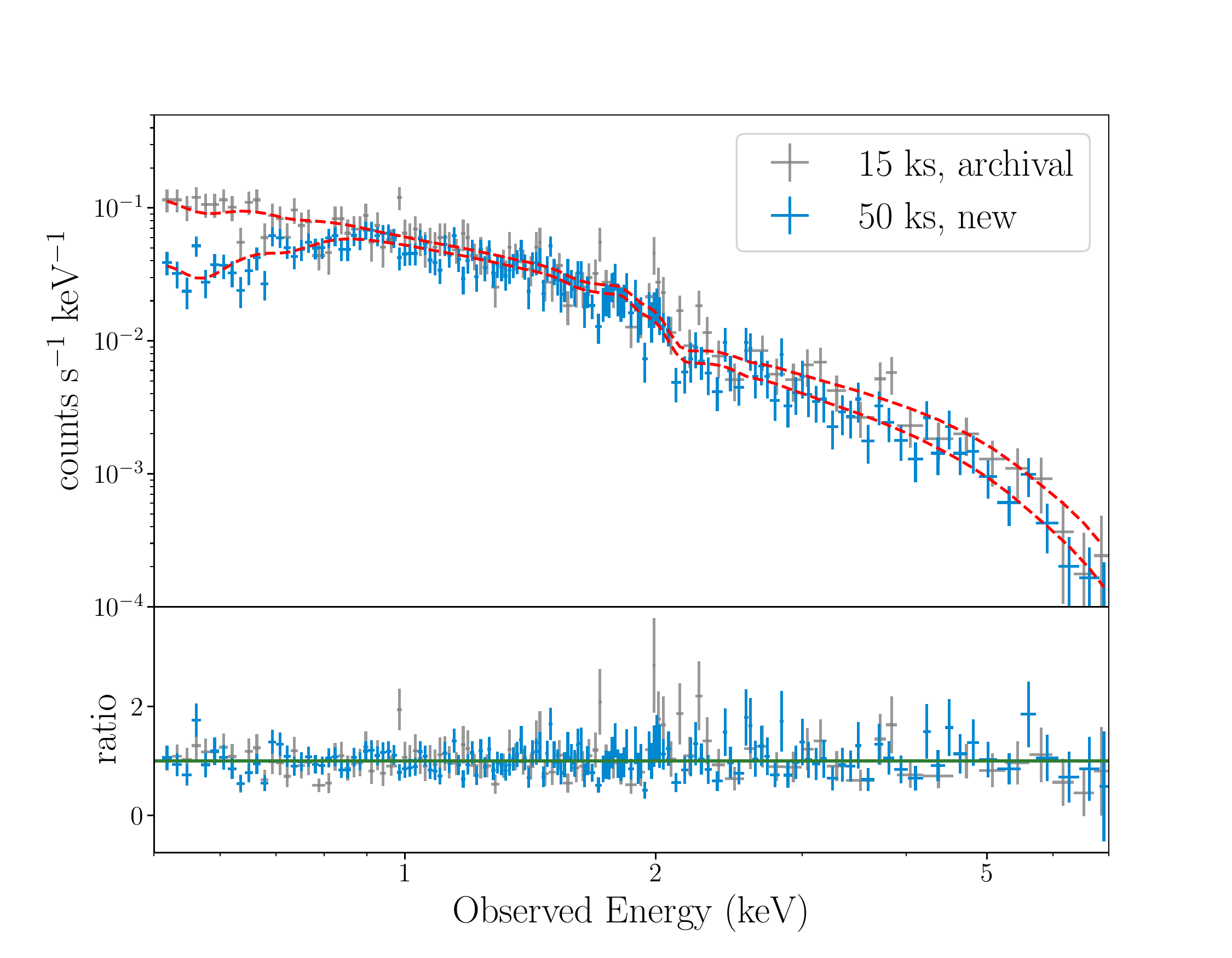}
\caption{\emph{Top:} The observed $0.5$--$7.0$ keV \emph{Chandra} spectrum of \IMBH{} is shown for both the 15 ks archival observation (grey points) and our new 50 ks observation (blue points), where the data have been folded through the instrument response. Both spectra appear to have a soft excess component, a feature seen in many narrow-line Seyfert 1 AGN.  We fit the spectrum with the model {\tt phabs$\times$zphabs$\times$(zpow+zbbody)}, fixing the Galactic absorption and redshift parameters at $N_{H} = 4.0 \times$ 10$^{20}$ cm$^{-2}$ and $z=0.14$. For each dataset, the best-fit models are shown in red. We list the best-fit values for each model in Section~\ref{sec:specfitting}, defined as the median of the distribution.  Because our analysis with \BAYMAX{} is restricted to the $2$--$7$ keV photons from \IMBH{}, our results are not affected by the soft emission component in the spectrum. In particular, although we detect variability between the two observations in the low-energy band, the $2$--$10$ keV fluxes are consistent with one another when we fit each spectra independently between $2$--$7$ keV with an absorbed power law.  \emph{Bottom:} Ratio of the data to the continuum model for \IMBH{}. The spectrum has been rebinned for plotting purposes.}
\label{fig:ChandraSpectrum}
\end{figure*}

 Both spectra have an excess of flux at soft X-ray energies ($<1$ keV) with respect to the power-law continuum, while the 15 ks data appear to catch the source in a higher flux state in the soft X-ray band (see Fig.~\ref{fig:ChandraSpectrum}).  Both of these behaviors are seen in AGN with a ``soft excess" component (e.g., \citealt{Lohfink2013, Lohfink2012}; see \citealt{Miniutti2009,Ludlam2015} for examples of soft X-ray excess in low-mass AGN candidates), an excess in emission above the extrapolated $2$--$10$ keV flux that is detected in over 50\% of Seyfert 1s \citep{Halpern1984,Turner&Pounds1989,Piconcelli2005,Bianchi2009,Scott2012}.  The physical origin of the soft excess remains uncertain; the shape is suggestive of a low-temperature high optical depth Comptonization of the inner accretion disk, however the temperature of this component appears to be constant over a wide range of black hole masses (and thus inferred accretion disk temperatures; see \citealt{Gierlinski&Done2004, Crummy2006}). The two most popular explanations for the soft excess are blurred ionized reflection from the inner parts of the accretion disk (e.g., \citealt{Fabian2002, Fabian2005, Gierlinski&Done2004, Crummy2006}) and Comptonization components (such as partial covering of the source by cold absorbing material, see \citealt{Boller2002, Tanaka2004}). 
\par Indeed, we find a statistically better fit when including an absorbed redshifted blackbody component to account for this soft excess ({\tt phabs$\times$zphabs$\times$(zpow + zbbody)}).  We fix the Galactic hydrogen column density (the photoelectric absorption component {\tt phabs}) to $4.21 \times 10^{20}$ cm$^{-2}$ \citep{Kalberla2005},  and the redshift to $z=0.14$.  In Figure~\ref{fig:ChandraSpectrum}, we show the X-ray spectrum of both observations, along with the best-fit XSPEC models.  For the archival dataset, we find best-fit values for intrinsic $N_{H} = 3.38^{+0.10}_{-0.10} \times 10^{20}$ cm$^{-2}$, power law component $\Gamma = 2.01^{+0.11}_{-0.12}$; and blackbody component $kT = 0.10^{+0.03}_{-0.05}$ keV.  For our new dataset, we find best-fit values for intrinsic $N_{H} = 4.07^{+0.10}_{-0.10} \times 10^{20}$ cm$^{-2}$, power law component $\Gamma = 2.51^{+0.10}_{-0.12}$; and blackbody component $kT = 0.11^{+0.05}_{-0.04}$ keV.  
\par However, because our analysis with \BAYMAX{} is restricted to the $2$-$7$ keV photons from \IMBH{}, our results are not affected by the soft emission component in the spectrum.  In particular, although we detect variability between the two observations in the low-energy band, the $2$--$10$ kev fluxes are consistent with one another (at the 99.7\% C.L.) when we fit each spectra independently between $2$--$7$ keV with an absorbed redshifted power law. For the 15 ks observations we calculate a total observed $2$--$10$ keV flux of $3.20^{+0.90}_{-0.80} \times 10^{-13}$ erg s$^{-1}$ cm$^{-2}$, while for the 50 ks observation we calculate a total observed $2$--$10$ keV flux of $2.23^{+1.0}_{-0.49} \times 10^{-13}$ erg s$^{-1}$ cm$^{-2}$ s$^{-1}$.  This corresponds to a rest-frame $2$--$10$ keV luminosity of $1.83^{+0.31}_{-0.40} \times 10^{43}$ erg s$^{-1}$ and $1.25^{+0.35}_{-0.21} \times 10^{43}$ erg s$^{-1}$ at $z=0.14$ (assuming isotropic emission). 
%

\section{Results}
\label{sec:results}
Analyzing the 15 ks \emph{Chandra} data with the EDSER option enabled, \IMBH{} appears to be an interesting dual AGN candidate. When binned, the data show a possible secondary source 0\farcs3 away from the primary (see Fig.~\ref{fig:ChandraImages}).  Although a possibly interesting result, classifying the source based on a qualitative analysis runs the risk of a false positive.  A statistical analysis is necessary before a discovery can be confirmed.  With an abundance of photons, and a robust model of the \emph{Chandra} PSF, in the following section we aim to unambigiously determine the true nature of \IMBH{}.  We first analyze each observation individually using \BAYMAX{}, and then combine the two (yielding a total of $n=723$ counts between $2$--$7$ keV).
\par We restrict our analysis to photons with (i) energies between $2$--$7$ keV and (ii) contained within a 4$\times$4 sky-pixel box (1.98\arcsec $\times$ 1.98\arcsec) centered on the nominal X-ray coordinates of the AGN.  This corresponds to $\sim$95\% of the encircled energy radius for the $2$--$7$ keV photons.  Because we expect $\lesssim$ 1 and 1.5 background counts within this region for the archival and new dataset, each photon is assumed to originate from either one ($M_{1}$) or two ($M_{2}$) point sources, with no background contamination.  The asymmetric PSF feature is within this extraction region, and sits approximately 0\farcs7 from the center of the AGN (see Fig.~\ref{fig:ChandraImages}). Within $2$-$7$ keV, there are $6$ and $12$ photons that spatially coincide with the feature for the 15 and 50 ks observations, respectively.  We mask the feature in both exposures before running \BAYMAX{}.
\par We run \BAYMAX{} with the initial conditions for the parameter vectors $\theta_{1}$ and $\theta_{2}$ as stated in Section~\ref{sec:methods}. When running \BAYMAX{} on our 15 ks and 50 ks observation individually, $k=1$ and thus we exclude the $\Delta x_{1}$ and $\Delta y_{1}$ from $\theta_{1}$ and $\theta_{2}$.  Further, we run \BAYMAX{} with the initializations for {\tt nestle} as described in Section~\ref{sec:methods}, with 500 active points and dlog$Z=0.1$
\par Our 15 ks observation has a total of $n=251$ counts between $2$--$7$ keV, while our 50 ks observation has a total of $n=472$ counts between $2$--$7$ keV. Using \BAYMAX{}, we calculate a Bayes factors (defined as the ratio of the evidence for the dual point source model to the single point source model) of $\frac{Z_{2}}{Z_{1}}=0.154$ and $\frac{Z_{2}}{Z_{1}}=0.102$ for the 15 ks and 50 ks observations, respectively.  This represents a Bayes factor of $\approx6.5$ and $\approx9.8$ in favor of a single point source.  The relative magnitudes of the $BF$ values are not surprising -- because the 15 ks observation has fewer counts than the 50 ks observation we expect there to be less evidence in favor of a given model.  Indeed, using the definitions presented in \cite{Kass&Raftery1995}, both of these $BF$ values are considered ``positive" against the dual point source model.  Further, the posterior distributions for $\theta_{2}$ are consistent across both datasets: the best-fit locations for $\mu_{P}$ and $\mu_{S}$ are consistent with one-another at the 95\% confidence interval, and the joint posterior distributions have shapes consistent with a single point source (consistent with the ``L" shape seen in Fig.~\ref{fig:BAYMAXsimulations}).
\par  Given that the individual analyses on each dataset favor the same model, and that we can treat the two spectra as the same between $2$--$7$ keV, we increase our statistical power and combine both datasets.  This yields a total of $n=723$ counts between $2$--$7$ keV.  Although we analyze the two observations jointly, we emphasize that the likelihoods for each observation are calculated independently of one another, and are a function of their respective PSF models.  Here, $k=2$ and $\Delta x_{1}$ and $\Delta y_{1}$ are included in parameter vectors for each model.   We use \BAYMAX{} to calculate a Bayes factor $\frac{Z_{2}}{Z_{1}}=7.40\times10^{-2}$. This represents a Bayes factor of $\approx$13.5 in favor of a single point source.  
\par To test the impact of the MCMC nature of nested sampling, we run \BAYMAX{} multiple times on the combined datasets.  We find consistent results, with a spread in ln$BF$-space well-described by a Gaussian distribution centered at ln$BF=2.6$ with standard deviation of 0.2. \emph{This Bayes factor strongly supports that the single point source model best describes the X-ray emission from SDSS J0914+0853.}  In Table~\ref{tab:Posteriors Single} we list the best-fit values (defined as the median value of the posterior distributions) for parameter vector $\theta_{1}$.
\par
\begin{figure*}
\centering
    \includegraphics[width=0.4\linewidth]{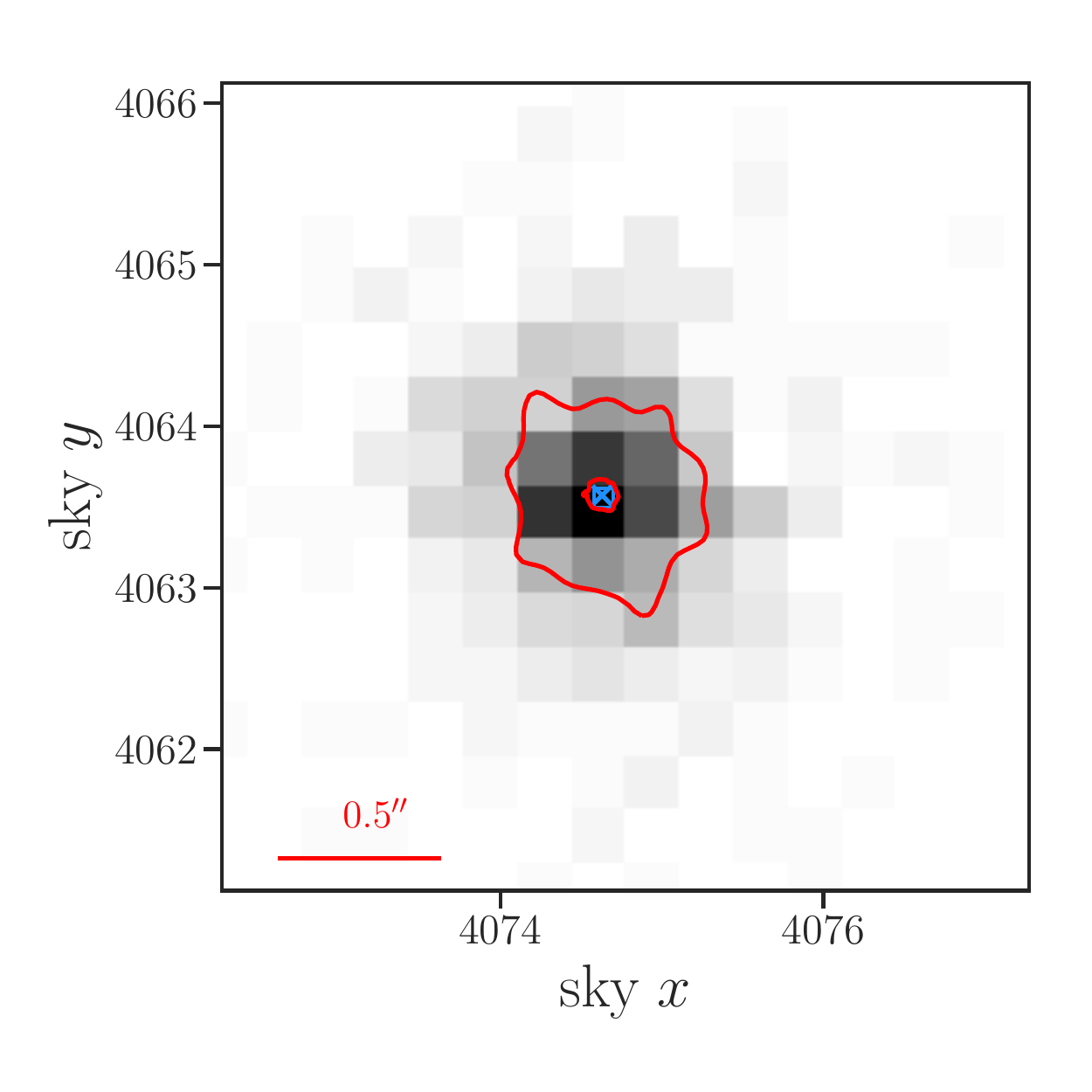}{\hskip 0pt plus 0.3fil minus 0pt}
    \includegraphics[width=0.44\linewidth]{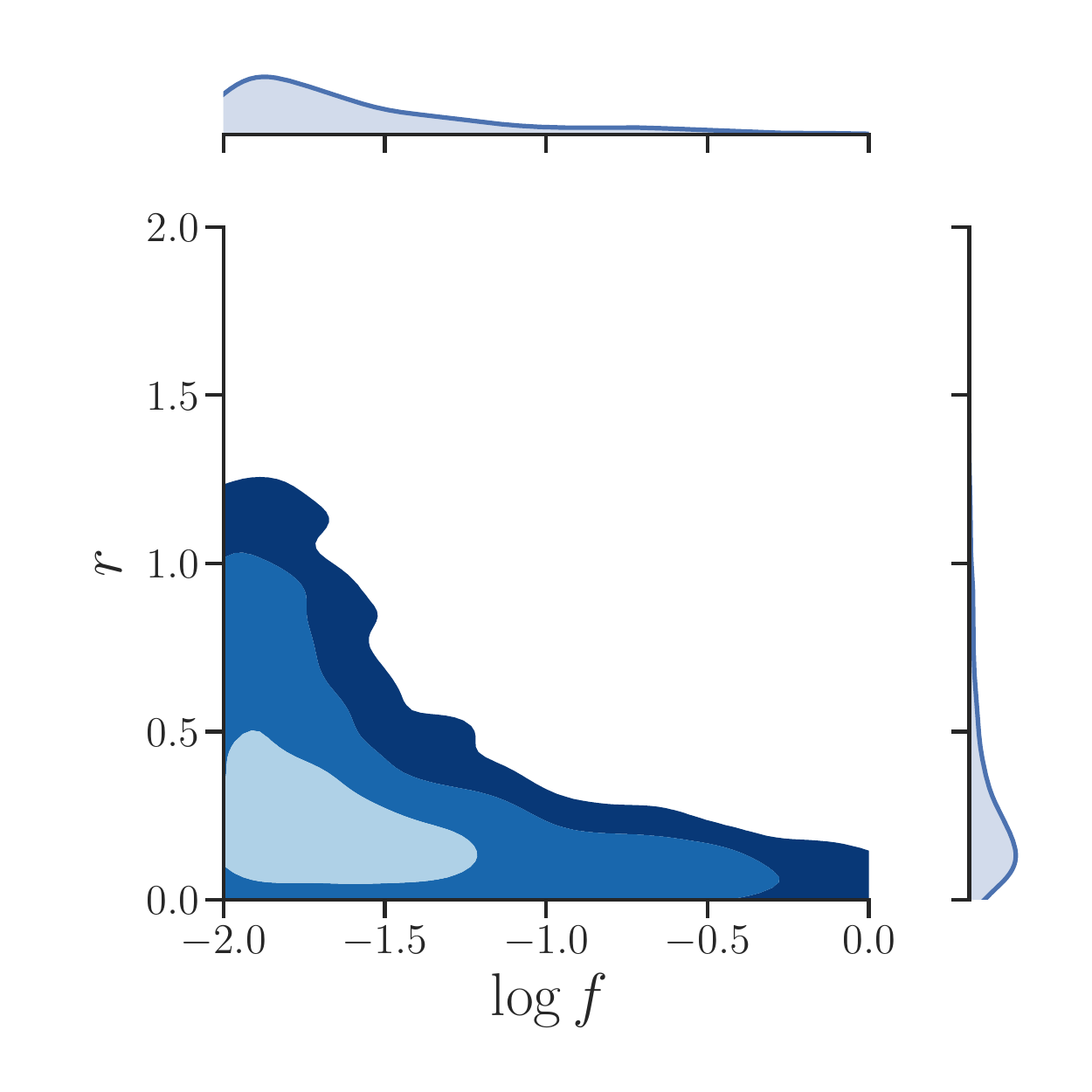}{\vskip 0.2cm plus 1fill}
    
\caption{\emph{Left}: The combined $2$--$7$ keV dataset (723 counts) for \IMBH{}, with the best-fit sky $x$ and sky $y$ positions for a primary ($\mu_{P}$, blue ``x") and secondary ($\mu_{S}$, blue square) AGN, and the respective 68\% and 95\% confidence intervals (red lines).  In order to more clearly see the results, we show a smaller area than shown in Fig.~\ref{fig:BAYMAXsimulations} (however, the binning of data remains the same). The spatial positions of the primary and secondary AGN are consistent with one-another.  \emph{Right}: Joint posterior distribution for the separation $r$ (in arcseconds) and the flux ratio (in units of $\log{f}$), with the marginal distributions shown along the border. 68\%, 95\%, and 99.7\% confidence intervals are shown in blue contours. At the 99.7\% confidence level, SDSS J091449 has a separation and flux ratio consistent with zero. We note that this particular joint-distribution shape is consistent with a single AGN, where at very large flux ratios the system is likely to have $r=0$, and at very large separations the system is likely to have $\log{f}=-2$.  }
\label{fig:BAYMAXresults}
\end{figure*}

 We examine the posterior distributions for $\theta_{2}$ to better understand our results. In Figure~\ref{fig:BAYMAXresults} we show the combined $2$--$7$ keV dataset ($\approx$ 65 ks, where the photons associated with the 15 ks exposure have been spatially shifted by the most-likely $\Delta x_{1}$ and $\Delta y_{1}$) with the best-fit $x$ and $y$ sky-coordinates for the primary and secondary AGN ($\mu_{P}$ and $\mu_{S}$), as well as the joint posterior distribution for the separation, $r$, and log flux ratio, $\log{f}$, parameters. Here, $r = \sqrt{ (\mu_{x,P}-\mu_{x,S})^{2} + (\mu_{y,P}-\mu_{y,S})^{2}}$.  Spatially, the best-fit locations for $\mu_{P}$ and $\mu_{S}$ are consistent with one-another at the 95\% confidence interval. Further, the joint posterior distribution has a shape consistent with a point source\,---\,the median values of the marginal posterior distributions are $r=0.15\pm0.5$ and $\log{f}=-1.6\pm0.4$, at very large flux ratios ($\log{f}\rightarrow{0}$) the separation is consistent with 0, and at very large separations ($r\rightarrow{2\arcsec}$) the flux ratio is consistent with 0.  The best-fit values for all the parameters in parameter vector $\theta_{2}$ are listed in Table~\ref{tab:Posteriors Single}.  
\par We investigate the influence of our prior distributions on our results. In particular, the Bayesian evidence automatically implements Occam's razor\,---\,the simpler model will be more easily favored than the more complicated one, unless the latter is significantly better at explaining the data. For our analysis, this means that the dual point source model needs enough data to overcome the inherent bias that \BAYMAX{} has towards favoring the single point source model.  Whenever the prior distribution is relatively broad compared with the likelihood function, the prior has fairly little influence on the posterior.  Thus, we re-run \BAYMAX{} with Gaussian prior distributions for $\mu$, $\mu_{P}$, and $\mu_{S}$:
\begin{equation}
    \mu = \mathcal{N}(\mu_{m}, \sigma^{2}),
\end{equation}
where $\mu_{m}$ and $\sigma^{2}$ represent the mean and variance of the distribution.  For $\mu_{P}$ and $\mu_{S}$, we set $\mu_{m}$ to the nominal X-ray positions of the potential primary and secondary AGN and set $\sigma$ to the observed separation between the two ($\sim0\farcs3$), given the 15 ks archival observation. For $\mu$ we set $\mu_{m}$ to the nominal X-ray position of the AGN, and similarly set $\sigma$ to 0\farcs3.  \BAYMAX{} calculates a Bayes factor of 10.8 $\pm$ 1.2, consistent within the errors of our previous measurement.  Further, the posterior distributions returned by \BAYMAX{} are consistent with those listed in Table~\ref{tab:Posteriors Single}.  We conclude that using sharper priors (comparable to the sharpness of our likelihoods), has no effect on our results. 
%
\begin{table}[t]
\begin{center}
\caption{Posterior Results for $\theta_{1}$ and $\theta_{2}$}
\label{tab:Posteriors Single}
\small
\begin{tabular*}{\columnwidth}{l@{\extracolsep{\fill}}r}
	\hline
	\hline
	\multicolumn{1}{c}{Parameter} & \multicolumn{1}{c}{Best-fit Value} \\
	\multicolumn{1}{c}{(1)} & \multicolumn{1}{c}{(2)} \\
	\hline
	\multicolumn{2}{c}{Single Point Source Model} \\
	\hline
	$\mu_{x}$	& 4074.6 $\pm$ 0.1	\\
    $\mu_{y}$	& 4063.6 $\pm$ 0.1	\\
    $\alpha$   & 138.70 \\
    $\delta$   & +8.89 \\
	$\Delta x_{1}$	& $-$14.8\arcsec $\pm$ 0\farcs{1}	\\
	$\Delta y_{1}$	& $-$25.8\arcsec $\pm$ 0\farcs{1}			\\
	$\Delta r_{1}$	& 29.8\arcsec $\pm$ 0\farcs{1}		\\
	\hline
	\multicolumn{2}{c}{Dual Point Source Model} \\
	\hline
	$\mu_{x, P}$	& 4074.6 $\pm$ 0.1	\\
    $\mu_{y, P}$	& 4063.6 $\pm$ 0.1	\\
	$\mu_{x, S}$	& 4074.6 $\pm$ 1.5	\\
    $\mu_{y, S}$	& 4063.4 $\pm$ 1.5	\\
    $\alpha_{P}$   & 138.70 \\
    $\delta_{P}$   &  +8.89 \\
    $\alpha_{S}$   & 138.70 \\
    $\delta_{S}$   & +8.89 \\
    $r$         & 0\farcs{15} $\pm$ 0\farcs{15}\\
    $\log{f}$    & $-$1.6 $\pm$ 0.4\\
	$\Delta x_{1}$	& $-$14.8\arcsec $\pm$ 0\farcs{1}	\\
	$\Delta y_{1}$	& $-$25.8\arcsec $\pm$ 0\farcs{1}			\\
	$\Delta r_{1}$	& 29.8\arcsec $\pm$ 0\farcs{1}		\\
	\hline
\end{tabular*}
\end{center}
Note. -- Columns: (1) Parameters from $\theta_{1}$: $\mu_{x}$ is the central x sky coordinate of the source, $\mu_{y}$ is the central y sky coordinate of the source, $\alpha$ is the central right ascension of the source in degrees, $\delta$ is the central declination of the source in degrees, $\Delta x_{1}$ is the translational astrometric shift in arcseconds, $\Delta y_{1}$ is the translational astrometric shift in arcseconds, and $\Delta r_{1}$ is the radial astrometric shift in arcseconds.  Parameters $\alpha$, $\delta$, and $\Delta r_{1}$ are not fit for by \BAYMAX{} but are calculated using $\mu_{x}$, $\mu_{y}$, $\Delta x_{1}$, and $\Delta y_{1}$. Parameters from $\theta_{2}$ are the same as $\theta_{1}$, where the underscore $P$ refers to the primary and $S$ refers to the secondary. Additionally: $r$ is separation between the two point sources in arcseconds and $\log{f}$ represents the log of the flux ratio; (2) the best-fit values from the Posterior Distributions, defined as the median of the distribution. All Posteriors distributions are unimodal, and thus the median is a good representation of the value with the highest likelihood. Error bars represent the 3$\sigma$ confidence level of each distribution.
\end{table}

\section{Discussion} 
\label{sec:discussion}
Our results support the hypothesis that the low-mass dual AGN candidate \IMBH{} is instead a single AGN.  Individually, we find $BF$ values of 6.5 and 9.8 in favor of a single point source for the 15 ks and 50 ks observations. When we combine the two datasets for a joint analysis, we find a $BF\sim13.5$ in favor of a single point source, and the posterior distributions are consistent with this model. Further, the prior distributions do not appear to have a great influence on our posteriors, reflecting that the data should be sufficient to favor the correct model, even when accounting for the Bayesian bias. In the following section we discuss the significance of our results by analyzing \BAYMAX{}'s capabilities across a range of parameter space for both the single and dual point source models. Assuming that \IMBH{} is indeed a dual AGN system, we investigate how the $BF$ determined by \BAYMAX{} depends on parameters $r$ and $f$.  In particular, we aim to understand where in parameter space \BAYMAX{} loses sensitivity for simulations with a comparable number of counts as our observations. %
\subsection{\BAYMAX{}'s Sensitivity Across Parameter Space}
\par The first step is to investigate how well \BAYMAX{} can classify a sample of simulated single AGN, i.e., our frequency of false-positives. This measurement will allow us to better define a range of Bayes factors that we can consider ``strongly" support the dual point source model. We simulate 100 single AGN via {\tt MARX}, assuming the same telescope configuration and spectrum as our new dataset.  Further, each simulation has 700 photons between $2$--$7$ keV.  We analyze each simulation with \BAYMAX{} and find that only $2$ are misclassified as a dual AGN with $BF>3$ (with the largest $BF=3.5$).  Thus, we define a $BF>3$ in favor of a dual AGN as ``strong evidence", while anything below this cut is classified as inconclusive. 
\par
\begin{figure*}
\centering
    \includegraphics[width=0.6\linewidth]{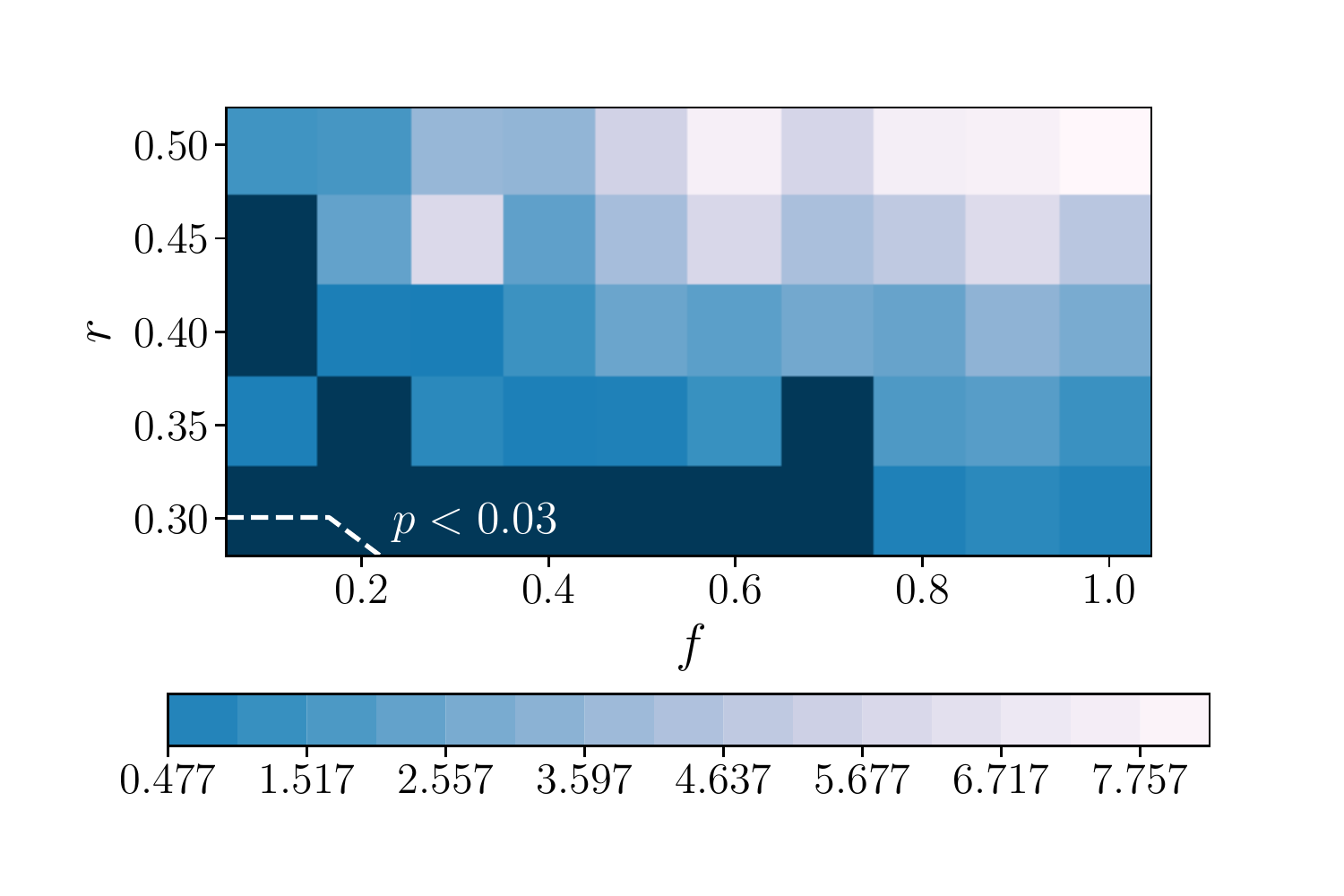}
    \caption{Bayes factor (defined as $Z_{2}/Z_{1}$) for simulated dual AGN with varying separation ($r$, in arcseconds) and flux ratios ($f$). For each point in parameter space we evaluated 100 simulations with randomized position angles (0-360$^{\circ}$) between the primary and secondary AGN.  Here we plot the logarithm of the mean $BF$ for each point in parameter space.  We enforce a cut of $BF>3$, where above this value the Bayes factor is classified as strongly in favor of the dual point source model. Points in parameter space with a $BF$ below this value are shown in dark blue. For a frequentist perspective, we add a contour (white dashed-line) where dual AGN with $f$ and $r$ values above the region have $<$3\% chance of being classified as a single AGN with $BF\ge13.5$, while all of parameter space has $<$5\%.  Assuming a null hypothesis that \IMBH{} is a dual AGN, we can reject the null hypothesis (with $p<0.03$) at $f>0.2$ for separations as low as 0\farcs3. }
\label{fig:SimulationGrid}
\end{figure*}

 We then run \BAYMAX{} on a suite of simulated dual AGN systems, generated via {\tt MARX}.  The simulations were created with the same assumptions as listed above.  Each simulation has 700 photons between $2$--$7$ keV, and each simulated AGN has the same $2$--$7$ keV spectrum as \IMBH{}, but with normalizations proportional to their flux ratio.  We simulated systems with separations that range between $0\farcs{3}$--$0\farcs{5}$ and flux ratios that range between $0.1$-$1.0$.  For each $r$--$f$ point in parameter space, we evaluate 100 simulations with randomized position angles between the primary and secondary.  Our results are shown in Figure~\ref{fig:SimulationGrid}, where we plot the logarithm of the mean $BF$ for each point in parameter space.  Consistent with expectations, \BAYMAX{} favors the dual point source model more strongly as the separation and flux ratio of a given dual AGN simulation increases, where we can expect $BF$ on the order of $\approx 10^{7}$ for systems with $r\ge0\farcs{5}$.  We enforce a cut of $BF>3$, where only $BF$ above this value are classified as strongly in favor of the dual point source model.  We find that we are sensitive to most flux ratios where $r\ge0\farcs{35}$, and for the smallest separations ($r\le0\farcs{35}$) \BAYMAX{} is capable of identifying the correct model when $f\ge0.8$
\subsection{A Quasi-Frequentist Approach}
Our analysis is intended to be a fully Bayesian inference, however some readers may find a frequenstist interpretation more intuitive. In the following section, we describe a potential interpretation of our results using a quasi-frequentist perspective.
\par On average, for separations below 0\farcs{35}, \BAYMAX{} will not necessarily favor the correct model for a dual AGN system.  For \IMBH{}, we estimate a possible separation of 0\farcs{3}, given the shallower \emph{Chandra} observation. However, the strength of the Bayes factor in favor of a single AGN has its own significance. From a frequentist perspective, we may ask what is the probability of measuring a $BF\ge13.5$ in favor of a single AGN \emph{if the system is dual AGN}.  In this specific scenario, our ``null hypothesis" is that \IMBH{} is a dual AGN system and our $p$-value represents the probability of measuring a $BF\ge13.5$ in favor of a single AGN.  Using our suite of dual AGN simulations, we analyze the probability of measuring a $BF\ge13.5$, as a function of $r$ and $f$.  Across all of parameter space, we find $p\le0.05$ and thus reject the null hypothesis at a 95\% confidence level. If we set our $p$-value threshold to $p<0.03$, we find that only for the smallest flux ratios ($f<0.2$) can the null hypothesis not be rejected for $r<0.3$ (see Fig.~\ref{fig:BAYMAXsimulations}).  Thus, the probability of \IMBH{} being a dual AGN system with a (1) flux ratio $f>0.3$, (2) separation $r>0\farcs{3}$, and (3) measured $BF=13.5$ in favor of a single AGN, is very low. \\ \\
We find that for observations with 700 counts \BAYMAX{} is sensitive to a large region of $r$--$f$ parameter space, such that if \IMBH{} were a dual AGN, we expect different results.  Our results and discussion highlight the importance of a robust, quantitative analysis of dual AGN candidates that are classified by their X-ray emission.  Most candidate dual AGN are discovered via indirect detection methods, such as narrow-line optical spectroscopy or optical/IR photometry.  However, directly detecting the X-ray emission unambigiously associated with a AGN is necessary for confirmation.  For candidate AGN with separations on the order of \emph{Chandra}'s resolution  ($<1\arcsec$), receiving observations with sufficient counts, paired with a robust model of the \emph{Chandra} PSF will allow for the most accurate analysis.  In particular, we may expect that most dual AGN candidates should have separations $<1\arcsec$, as at a distance of 200 Mpc ($z\approx0.05$) the physical-to-angular scale becomes 1.0 kpc/arcsec.  Given the small number of currently confirmed dual and binary AGN, tools such as \BAYMAX{} will be important for a precise measurement of the dual AGN rate, and as a result, an improved physical understanding of the evolution of SMBHs and their activity.

\section{Conclusions}
In this work, we present the first analysis by \BAYMAX{}, a tool that uses a Bayesian framework to statistically and quantitatively determine whether a given observation is best described by one or two point sources. \BAYMAX{} takes calibrated events from a \emph{Chandra} observation and compares them to simulations based on single and dual point source models.  \BAYMAX{} determines the most likely model by the calculation of the Bayes factor, which represents the ratio of the plausibility of the observed data $D$, given the model $M_{j}$ and parameterized by the priors. We present the results of \BAYMAX{} analyzing the lowest-mass dual AGN candidate \IMBH{}, which was originally targeted as a dual AGN based on shallow archival \emph{Chandra} imaging.  The 15 ks exposure appears to have a secondary AGN $\sim$ 0\farcs{3} from a primary AGN  We received a new 50 ks \emph{Chandra} exposure, with (i) a shorter frame time to avoid pileup and (ii) a different roll angle, with the aim of unambiguously determining the true accretion nature of the AGN. The main results and implications of this work can be summarized as follows:
\begin{enumerate}
    \item Analyzing our new 50 ks observation, we find (by visual analysis) that the $2$--$7$ keV emission more closely resembles that of a single point source.  Both spectra have an excess of flux at soft X-ray energies ($<1$ keV) with respect to the power-law continuum, while the 15 ks observation appear to catch the source in a higher flux state in the soft X-ray band.  Both of these behaviors are seen in AGN with a ``soft excess" component, and we fit our spectra with an absorbed redshifted powerlaw and blackbody ({\tt phabs$\times$zphabs$\times$(zpow + zbbody)}).  For the archival dataset, we find best-fit values for intrinsic $N_{H} = 3.38^{+0.10}_{-0.10} \times 10^{20}$ cm$^{-2}$, power law component $\Gamma = 2.01^{+0.11}_{-0.12}$; and blackbody component $kT = 0.10^{+0.03}_{-0.05}$ keV.  For our new dataset, we find best-fit values for intrinsic $N_{H} = 4.07^{+0.10}_{-0.10} \times 10^{20}$ cm$^{-2}$, power law component $\Gamma = 2.51^{+0.10}_{-0.12}$; and blackbody component $kT = 0.11^{+0.05}_{-0.04}$ keV.  
    \item We find that the $2$--$7$ keV emission is consistent between the two observations, and fit the spectra in this energy-range with an absorbed redshifted powerlaw. For the 15 ks observations we calculate a total observed $2$--$10$ keV flux of $3.20^{+0.90}_{-0.80} \times 10^{-13}$ erg s$^{-1}$ cm$^{-2}$, while for the 50 ks observation we calculate a total observed $2$--$10$ keV flux of $2.23^{+1.0}_{-0.49} \times 10^{-13}$ erg s$^{-1}$ cm$^{-2}$ s$^{-1}$.  This corresponds to a rest-frame $2$--$10$ keV luminosity of $1.83^{+0.31}_{-0.40} \times 10^{43}$ erg s$^{-1}$ and $1.25^{+0.35}_{-0.21} \times 10^{43}$ erg s$^{-1}$ at $z=0.14$ (assuming isotropic emission). 
    \item We use \BAYMAX{} to analyze the 15 ks and 50 ks observations both individually, as well as combined, restricting our analysis to photons with energies between $2$--$7$ keV. Using \BAYMAX{} we calculate a Bayes factor in favor of the single point source model of $\approx$ 6.5 and 9.8 for the 15 ks and 50 ks observations, respectively. When combining the two observations, we calculate a Bayes factor of 13.5 in favor of a single point souce.  To test the impact of the MCMC nature of nested sampling, we run \BAYMAX{} multiple times on the combined datasets.  We find consistent results, with a spread in ln$BF$-space well-described by a Gaussian distribution centered at ln$BF=2.6$ with standard deviation of 0.2.
    \item Our posterior distributions for both the single and dual point source model further support that \IMBH{} is a single AGN. Spatially, the best-fit locations from $\mu_{P}$ and $\mu_{S}$ are consistent with one-another within the 68\% error level. Further, the joint posterior distribution has a shape expected from a single point source\,---\,the median values of the marginal posterior distributions are $r=0.15\pm0.5$ and $\log{f}=-1.6\pm0.4$, at very large flux ratios ($\log{f}\rightarrow{0}$) the separation is consistent with 0, and at very large separations ($r\rightarrow{2\arcsec}$) the flux ratio is consistent with 0.
    \item We investigate the influence of our prior distributions, by running \BAYMAX{} with Gaussian prior distributions for $\mu$, $\mu_{P}$, $\mu_{S}$. \BAYMAX{} calculates a Bayes factor in favor of a single point source of $10.8\pm1.2$, consistent within the errors of our initial measurement. Further, the posterior distributions returned by \BAYMAX{} are consistent with those listed in Table~\ref{tab:Posteriors Single}. 
    \item We investigate how the Bayes factor determined by \BAYMAX{} depends on the separation and flux ratio of a given dual AGN system.  We find that for \emph{Chandra} observations with at least 700 counts between $2$--$7$ keV, \BAYMAX{} is capable of strongly favoring the correct model for most flux ratios when $r\ge0\farcs{35}$.  For the smallest separations ($r\le0\farcs{3}$), \BAYMAX{} is capable of identifying the correct model when the flux ratio $f\ge0.8$.
    \item From a quasi-frequentist perspective, we estimate the probability of measuring a $BF\ge13.5$ in favor a single AGN, using a null hypothesis that \IMBH{} is actually a dual AGN. Across all of parameter space ($0\farcs{3}<r<0\farcs{5}$ and $0.1<f<1.0$), we find $p\le0.05$ and can reject the null hypothesis at a 95\% confidence level. Thus, the probability of \IMBH{} being a dual AGN system with a (1) flux ratio $f>0.3$, (2) separation $r>0\farcs{3}$, and (3) measured $BF=13.5$ in favor of a single AGN, is very low.
\end{enumerate}
We have shown through various analyses that there is an absence of evidence supporting \IMBH{} as a dual AGN system. Specifically, \BAYMAX{} estimates a Bayes factor strongly in favor of a single AGN and posterior distributions for a possible separation and flux ratio between a primary and secondary AGN are consistent with 0.  Moving forward, statistical analyses with \BAYMAX{} on \emph{Chandra} observations of dual AGN candidates will be important for a robust measurement of the dual AGN rate across our visible universe. Lastly, our Bayesian framework will eventually be capable for more general analyses, such as evaluating binary active star systems.\\

A.F. and K.G. acknowledge support provided by the National Aeronautics and Space Administration through \emph{Chandra} Award Number TM8-19007X issued by the \emph{Chandra} X-ray Observatory Center, which is operated by the Smithsonian Astrophysical Observatory for and on behalf of the National Aeronautics Space Administration under contract NAS8-03060. We also acknowledge support provided by the National Aeronautics and Space Administration through \emph{Chandra} Award Number GO7-18087X issued by the \emph{Chandra} X-ray Observatory Center, which is operated by the Smithsonian Astrophysical Observatory for and on behalf of the National Aeronautics Space Administration under contract NAS8-03060. A.F. thanks Abderahmen Zoghbi for helpful discussion regarding {\tt PyMC3}. This research has made use of NASA's Astrophysics Data System.

\software{ {\tt CIAO} (v4.8; \citealt{Fruscione2006}), \\
XSPEC (v12.9.0; \citealt{Arnaud1996}), \\
{\tt nestle} (https://github.com/kbarbary/nestle), \\
{\tt PyMC3} \citep{Salvatier2016}, \\ 
{\tt SAOTrace} (http://cxc.harvard.edu/cal/Hrma/Raytrace/SAOTrace.html), \\
{\tt MARX} (v5.3.3; \citealt{Davis2012})}

\bibliographystyle{aasjournal}
\bibliography{foord.bib}

\end{document}